\pdfoutput=1
\RequirePackage{fix-cm}

\documentclass[twocolumn,epjc3]{svjour3}
\RequirePackage[colorlinks,citecolor=blue,urlcolor=blue,linkcolor=blue]{hyperref}
\RequirePackage[T1]{fontenc}
\smartqed  

\RequirePackage{graphicx}
\RequirePackage{multirow}
\RequirePackage{mathptmx}      
\RequirePackage{latexsym}
\RequirePackage{amsmath}
\RequirePackage{nicefrac}
\RequirePackage{amssymb}
\RequirePackage{multirow}
\RequirePackage[sort&compress,numbers]{natbib}
\RequirePackage{mdwlist}
\RequirePackage{lineno}

\RequirePackage[english]{babel}
\journalname{Eur. Phys. J. C}

\begin{document}

\title{Search for non-relativistic Magnetic Monopoles with IceCube}

\onecolumn

\author{IceCube Collaboration: M.~G.~Aartsen\thanksref{Adelaide}
\and R.~Abbasi\thanksref{MadisonPAC}
\and M.~Ackermann\thanksref{Zeuthen}
\and J.~Adams\thanksref{Christchurch}
\and J.~A.~Aguilar\thanksref{Geneva}
\and M.~Ahlers\thanksref{MadisonPAC}
\and D.~Altmann\thanksref{Erlangen}
\and C.~Arguelles\thanksref{MadisonPAC}
\and T.~C.~Arlen\thanksref{PennPhys}
\and J.~Auffenberg\thanksref{Aachen}
% ...
}
\authorrunning{IceCube Collaboration}
\thankstext{email_schoenen}{Corresponding author: schoenen@physik.rwth-aachen.de}
\thankstext{email_lotfi}{Corresponding author: mohamed.lotfi.benabderrahmane@desy.de}
\thankstext{a}{Physics Department, South Dakota School of Mines and Technology, Rapid City, SD 57701, USA}
\thankstext{b}{Earthquake Research Institute, University of Tokyo, Bunkyo, Tokyo 113-0032, Japan}
\thankstext{c}{NASA Goddard Space Flight Center, Greenbelt, MD 20771, USA}
\institute{III. Physikalisches Institut, RWTH Aachen University, D-52056 Aachen, Germany \label{Aachen}
\and School of Chemistry \& Physics, University of Adelaide, Adelaide SA, 5005 Australia \label{Adelaide}
% ...
\and Dept.~of Physics and Astronomy, University of Canterbury, Private Bag 4800, Christchurch, New Zealand \label{Christchurch}
% ...
\and Erlangen Centre for Astroparticle Physics, Friedrich-Alexander-Universit\"at Erlangen-N\"urnberg, D-91058 Erlangen, Germany \label{Erlangen}
\and D\'epartement de physique nucl\'eaire et corpusculaire, Universit\'e de Gen\`eve, CH-1211 Gen\`eve, Switzerland \label{Geneva}
% ...
\and Dept.~of Physics and Wisconsin IceCube Particle Astrophysics Center, University of Wisconsin, Madison, WI 53706, USA \label{MadisonPAC}
% ...
\and Dept.~of Physics, Pennsylvania State University, University Park, PA 16802, USA \label{PennPhys}
% ...
\and DESY, D-15735 Zeuthen, Germany \label{Zeuthen}
} 

\date{Received: \today / Accepted: date}

\maketitle
\twocolumn

\begin{abstract}
The IceCube Neutrino Observatory is a large Cherenkov detector instrumenting $1\,\mathrm{km}^3$ of Antarctic ice.
The detector can be used to search for signatures of particle physics beyond the Standard Model.
Here, we describe the search for non-relativistic, magnetic monopoles as remnants of the GUT (\textbf{G}rand \textbf{U}nified \textbf{T}heory) era shortly after the Big Bang.
Depending on the underlying gauge group these monopoles may catalyze the decay of nucleons via the Rubakov-Callan effect with a cross section suggested to be in the range of $10^{-27}\,\mathrm{cm^2}$ to $10^{-21}\,\mathrm{cm^2}$.
In IceCube, the Cherenkov light from nucleon decays
along the monopole trajectory would produce a characteristic hit pattern.
This paper presents the results of an analysis of first data taken from May 2011 until May 2012 with a dedicated slow-particle trigger for DeepCore, a subdetector of IceCube.
A second analysis provides better sensitivity
for the brightest non-relativistic monopoles using data taken from May 2009 until May 2010.
In both analyses no monopole signal was observed.
For catalysis cross sections of $10^{-22}\,(10^{-24})\,\mathrm{cm^2}$ the flux of non-relativistic GUT monopoles
is constrained up to a level of $\Phi_{90} \le 10^{-18}\,(10^{-17})\,\mathrm{cm^{-2}s^{-1}sr^{-1}}$ at a 90\% confidence level,
which is three orders of magnitude below the Parker bound.
The limits assume a dominant decay of the proton into a positron and a neutral pion.
These results improve the current best experimental limits by one to two orders of magnitude, for a wide range of assumed speeds and catalysis cross sections.

\end{abstract}

\section{Introduction}\label{sec:intro}

Magnetic monopoles are particles carrying a quantized magnetic charge and are predicted in various theories.
In classical electrodynamics, their existence would symmetrize Maxwell's equations with respect to the sources of the electromagnetic field.
Quantum mechanically, the existence of magnetic monopoles implies that both electric charge and the hypothetical magnetic charge, are quantized, given that the associated electromagnetic fields still satisfy Maxwell's equations \cite{Dirac:1931}.
The resulting magnetic elementary charge, called the Dirac charge $g_{\mathrm{D}}$, is

\begin{equation}
	g_{\mathrm{D}}=\frac{e}{2\alpha},
\end{equation}
where $e$ is the electric elementary charge and $\alpha$ is the fine structure constant.

In Grand Unified Theories (GUTs) \cite{Georgi:1974FirstGUT} magnetic monopoles appear as stable, finite energy solutions of the field equations \cite{tHooft:1974MonopoleSolution,Polyakov:1974MonopoleSolution}.
The predicted masses range from $10^5\,\mathrm{GeV}$ to $10^{17}\,\mathrm{GeV}$ 
\cite{Georgi:1974_Interaction_GUT,Daniel:1979_SU5Monopoles,Lazarides,Kephart,Wick:CosmicFluxMM} and their magnetic charges are integer multiples of the Dirac charge  $	g_{\mathrm{D}} $. 
The lower part of the mass range up to $\sim 10^{13}\,\mathrm{GeV}$ refers to intermediate mass monopoles (IMMs) which arise from intermediate stages of symmetry breaking below the GUT scale.
In contrast the superheavy monopoles with masses at the GUT scale may have been created during the phase transition 
associated with the spontaneous breakdown of the unified gauge symmetry in the early universe at $\sim 10^{-36}\,\mathrm{s}$ after the Big Bang \cite{Kibble}.
The monopole mass and charge depend on the underlying gauge group, the symmetry breaking hierarchy, and the type and temperature of the phase transition in a particular GUT.

Since magnetic monopoles are stable, they should still be present in cosmic rays. The number density today depends on the existence of an inflationary epoch and on the time of creation, which could be before, during or after this epoch \cite{Preskill:MM}. Since then, monopoles have been accelerated by large-scale cosmic magnetic fields. The kinetic-energy gain by passing through a magnetic field $B$ is given by 

\begin{equation}
	E_{\mathrm{kin}}=g\int\limits_{\mathrm{path}} \vec{B} \cdot d\vec{l},
\end{equation}
where $g =n\cdot g_{\mathrm{D}} $ is the magnetic charge. The maximum kinetic energy of a magnetic monopole due to acceleration in cosmic magnetic fields 
is rather uncertain but can reach $\sim 10^{14}\,\mathrm{GeV}$ \cite{Wick:CosmicFluxMM}. Therefore, monopoles with masses at, or above, this energy scale should be non-relativistic. Based on the propagation of magnetic monopoles in the Galactic magnetic field an upper bound on the monopole flux can be calculated, assuming the Galactic magnetic field does not decrease faster than it can be regenerated. This assumption constrains the monopole flux to be less than $10^{-15}\,\mathrm{cm}^{-2}\,\mathrm{s}^{-1}\,\mathrm{sr}^{-1}$, which is called the Parker Bound \cite{Turner:ParkerBound,Groom:SearchSupermassiveMM}. Taking into account the fields during galaxy formation, the limit was extended by Adams et al. to be less than $10^{-16}\,\mathrm{cm}^{-2}\,\mathrm{s}^{-1}\,\mathrm{sr}^{-1}$ for monopoles with masses below $10^{17}\,\mathrm{GeV}$ \cite{ExtensionParkerBound}.

Many experiments have searched for relic magnetic monopoles, but there is no experimental proof for their existence.
The current best limits for magnetic monopoles constrain their flux to a level of $\sim 10^{-16}-10^{-18}\,\mathrm{cm}^{-2}\,\mathrm{s}^{-1}\,\mathrm{sr}^{-1}$
depending on the monopole speed and interaction mechanism \cite{FinalMACRO,MACRO:SlowMonopoles,icecube:relMPs,IceCube:ICRC2013_relMPs}.
Consequently, searches for magnetic monopoles require very large detectors.

The IceCube Neutrino Observatory currently is the world's largest neutrino detector.
The primary goal is the detection of Cherenkov light from electrically-charged secondary particles produced when high-energy astrophysical neutrinos interact in the surrounding matter \cite{IceCubeSensitivity}.
However, IceCube can also be used to search for magnetic monopoles. Depending on their speed monopoles have different 
signatures in  IceCube. Relativistic monopoles with a speed above the Cherenkov threshold, e.g. $\beta \approx 0.76$ in ice, can be detected by the Cherenkov light they directly produce \cite{Tompkins:CherenkovEmission}. Non-relativistic monopoles that catalyze the decay of nucleons in the detector medium can, in contrast, be detected by the Cherenkov light from electrically charged secondary particles produced in subsequent nucleon decays along the monopole trajectory (Sec. \ref{subsec:Rubakov-Callan_Effect}). Therefore, different analysis strategies are needed in order to cover both detection channels.
This paper presents the results of a search for non-relativistic magnetic monopoles which would catalyze the proton decays via the Rubakov-Callan effect in IceCube.

\section{Monopole Detection with IceCube}
\subsection{The IceCube Detector \label{icecube}}
The IceCube Neutrino Observatory consists of the in-ice detector, IceCube, and the surface air shower detector, IceTop.
It is located at the geographic South Pole.
For the in-ice detector $1\,\mathrm{km}^3$ of the Antartic ice, which is used as detection medium, has been instrumented.
The detector consists of 86 strings 
equipped with 60 digital optical modules (DOMs) each.
The DOM, the sensor of the IceCube detector, consists of a glass pressure housing enclosing a $25.4\,\mathrm{cm}$ diameter Hamamatsu photomultiplier tube (PMT) with the electronics needed for signal digitization, and a set of LEDs for calibration purposes \cite{IceCube:PMT,icecube:daq}.
Signals that pass a threshold of about $0.25$ photo-electrons are  digitized and recorded. This process is called a DOM launch or for simplicity a \emph{hit} in the following. Two hits are labeled as hard local coincidences (also called HLC pair), if their time difference is less than $1\,\mathrm{\mu s}$ and the corresponding DOMs are nearest or next-to-nearest neighbors on the same string.
The recorded data is sent to the surface and a trigger 
algorithm evaluates the time and position of the hits and
decides whether they form an event. For example, for relativistic particles a simple multiplicity trigger requiring eight HLC hits within a time window of $5\,\mathrm{\mu s}$, called SMT8, is used.
The DOMs are deployed at depths between $1450\,\mathrm{m}$ and $2450\,\mathrm{m}$ \cite{IceCubePerform}.
At depths below $2100\,\mathrm{m}$, eight inner strings
are placed with smaller separations from each other and thus form a region of denser instrumentation. 
Together with seven central standard strings 
they form DeepCore, a low-energy sub-detector \cite{IceCubeDeepCore}.
The construction of IceCube was completed December 16, 2010 but data taken during intermediate construction stages were already used for physics analyses during earlier years. 
One of the two presented analyses uses data taken from May 2009 to May 2010, when IceCube was operating 
in its 59-string configuration (IC-59). The other analysis uses the fully installed detector.

\subsection{The Rubakov-Callan Effect}\label{subsec:Rubakov-Callan_Effect}

Non-relativistic magnetic monopoles would themselves be too
slow to emit Cherenkov light when propagating through the IceCube detector.
However, relativistic charged secondary particles, produced in monopole interactions with the surrounding matter,
can produce Cherenkov light and thus can be detected by the IceCube detector.

The energy loss of a magnetic monopole due to ionization can be described by a modified Bethe-Bloch formula \cite{Ahlen:modified_BetheBloch,Kazama:BetheBloch_Correction,Bloch:Correction}, which is valid for speeds $\beta > 0.1$. For lower speeds in the range from $\beta = 10^{-3}$ to $10^{-2}$ Ahlen and Kinoshita introduced a model to calculate the energy loss of magnetic monopoles \cite{Ahlen:IonizationSlowMP}. Later, Ritson extended this model for speeds below $\beta = 10^{-3}$ \cite{Ritson:IonizationSlowMP}. For magnetic monopoles with e.g. $\beta = 10^{-3}$ the energy loss is of the order of $20\,\mathrm{MeV}\,\mathrm{g^{-1}}\,\mathrm{cm^{2}}$ \cite{Groom:SearchSupermassiveMM}. Only electrons above the Cherenkov threshold of $\sim 0.28\,\mathrm{MeV}$ kinetic energy emit detectable 
Cherenkov light. 
However, the maximum transferred energy of a monopole with e.g. $\beta = 10^{-3}$ on an atomic electron 
is typically $E_{\mathrm{max}} \simeq 10\,\mathrm{eV}$ 
and no Cherenkov light is produced.

\begin{figure}
	\centering
	\includegraphics[width=0.9\columnwidth]{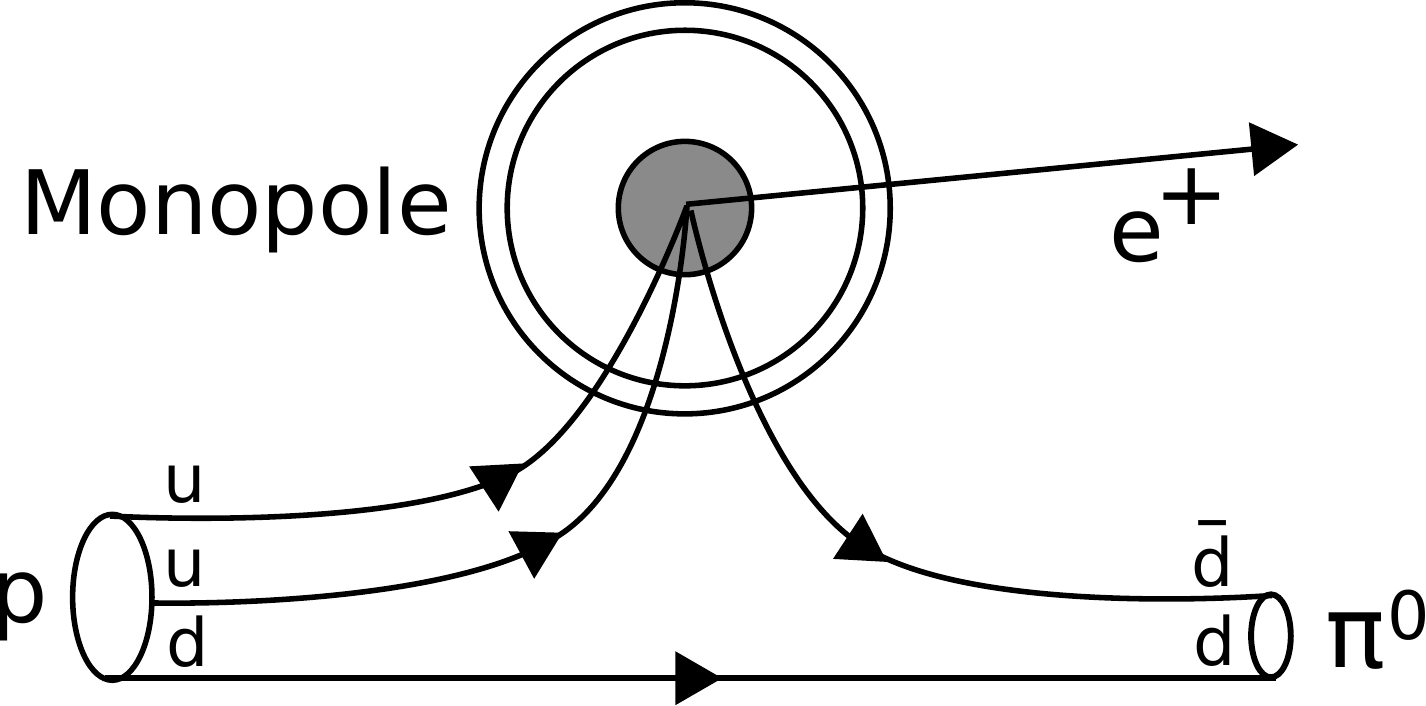}
	\caption{Illustration of a proton decay into a positron and a neutral pion catalyzed by a GUT monopole.}
	\label{fig:protondecay}
\end{figure}

Rubakov \cite{Rubakov:protondecay} and Callan \cite{CallanJr:protondecay} showed that some monopoles could catalyze nucleon decays along their trajectories (Rubakov-Callan effect).
This effect depends on the gauge group of the respective GUT theory \cite{DawsonPRD27:GUTdependence,Walsh:GUTdependence}
and on assumptions, e.g. on the fermion masses or the relative velocity between the quarks and the monopole, used in the calculation \cite{DawsonPRD28:GUTdependence,Rubakov:GUTdependence}.
Furthermore, this process is not possible for intermediate mass monopoles with masses below $\sim 10^{13}\,\mathrm{GeV}$ \cite{Kephart}.
Therefore, the sensitivity of this analysis is contrained to heavier monopoles (GUT scale).
Figure \ref{fig:protondecay} illustrates the catalyzed decay of a proton by a GUT monopole into a positron and a neutral pion:

\begin{equation}
	M + p \rightarrow M + e^+ + \pi^0.
	\label{eqn:promising_decay_channel}
\end{equation}
For this decay channel almost the full rest mass energy of the proton is transferred to electromagnetic particles.
Because of the high light yield this channel is used as a benchmark in the analyses.

The catalysis cross section for nucleon decays $\sigma_{\mathrm{cat}}$ depends
not only on the cross section $\sigma_0$ \cite{Rubakov:ValuesSigma0}, but also on the monopole speed $\beta=v/c$:

\begin{eqnarray}
	\sigma_{\mathrm{cat}} = \begin{cases}
                   			\frac{\sigma_0}{\beta} &\text{ for }\beta \geq \beta_0\\
                   			\frac{\sigma_0}{\beta} \cdot F(\beta) &\text{ for }\beta < \beta_0.
               			\end{cases}
    \label{eqn:sigma_cat_theo}
\end{eqnarray}
The correction $F(\beta) = \left( \frac{\beta}{\beta_0} \right)^\gamma$ 
takes into account an additional angular momentum of the 
monopole-nucleus-system 
and becomes  relevant for speeds below the speed threshold $\beta_0$.
Depending on the sign of $\gamma$ the catalysis cross section is enhanced or suppressed.
Both parameters $\gamma$ and $\beta_0$ depend on the nucleus \cite{Arafune:RubakovVelocityDependent}.
Current estimates for the catalysis cross sections are of the order of $10^{-27}\,\mathrm{cm}^2$ to $10^{-21}\,\mathrm{cm}^2$ \cite{Nath:sigma_cat}.

\begin{figure}
	\centering
	\includegraphics[width=0.9\columnwidth]{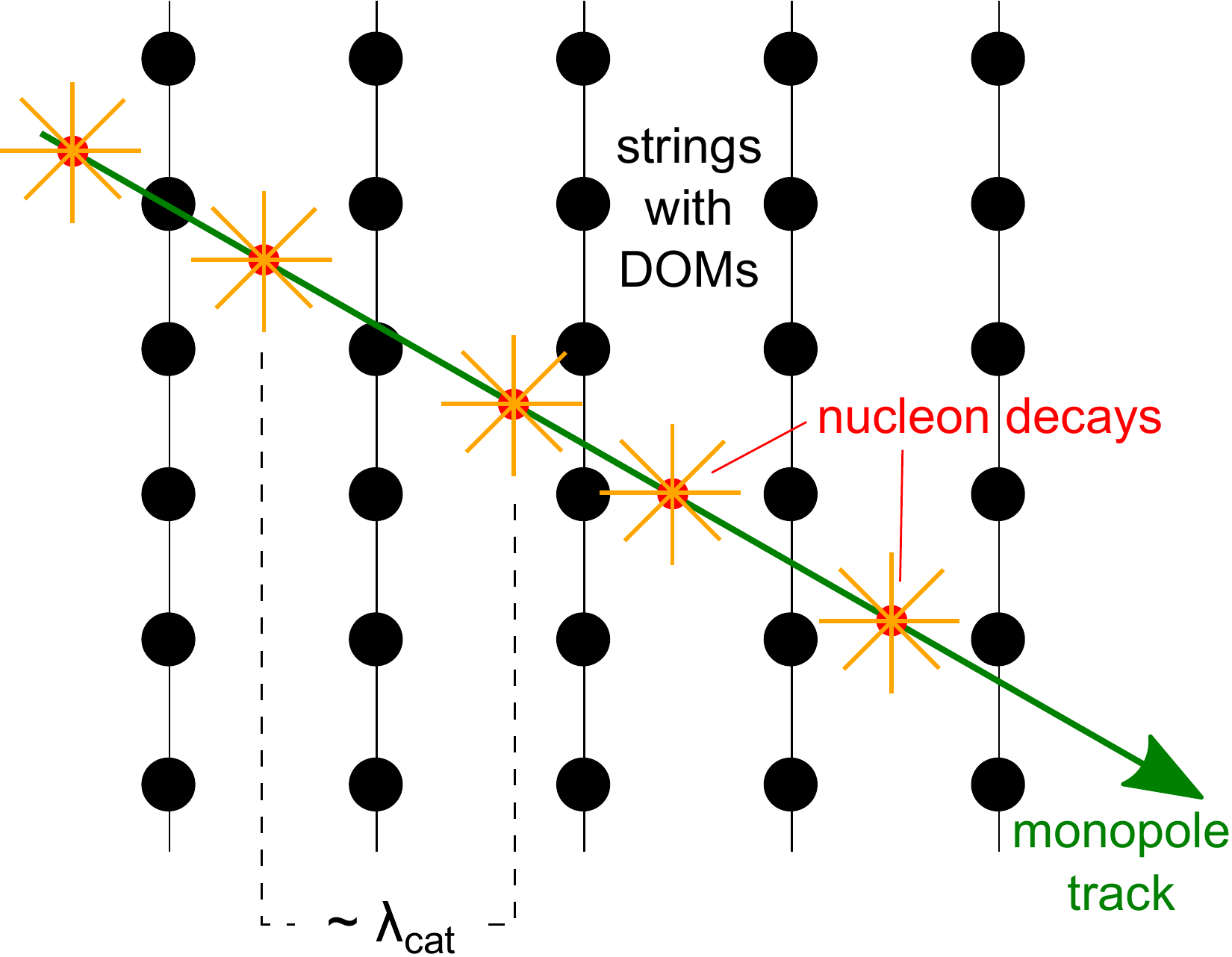}
	\caption{Illustration of the signature of a non-relativistic magnetic monopole (green) catalyzing nucleon decays (red) along its track in IceCube. The resulting cascades with mean distances $\lambda_{\mathrm{cat}}$ are symbolized by orange rays.}
	\label{fig:monopole_signature}
\end{figure}

The Rubakov-Callan effect results in
 small  electromagnetic or hadronic cascades from catalyzed nucleon decays 
along the monopole trajetory through the detector. 
This is illustrated in Fig. \ref{fig:monopole_signature}.
Experimentally, the relevant parameter is the mean free path $\lambda_{\mathrm{cat}}$ between two  decays. That is

\begin{equation}
	\lambda_{\mathrm{cat}} = \frac{1}{\sigma_{\mathrm{cat}} \cdot n},
	\label{eqn:mean_free_path}
\end{equation}
where $n$ is the particle density of the medium through which the monopole propagates.

The energy of each cascade, and therefore the number of emitted Cherenkov photons, depends on the decay channel (e.g. Eq. \ref{eqn:promising_decay_channel}). 
A general quantity is the track length, $l_{\gamma}$, that a relativistic particle carrying a single electric charge would have to travel in order to emit the same number of Cherenkov photons as the average number expected from a proton decay, $N_{\gamma}$ \cite{KowalskiCherenkovYield}.
Using this track length per proton decay, $l_{\gamma}$, the monopole's mean free path $\lambda_{\mathrm{cat}}$ can be converted into the light yield per monopole track length $\hat{l}$.

\begin{equation}
    \hat{l} = \frac{l_{\gamma}}{\lambda_{\mathrm{cat}}} \propto \frac{N_{\gamma}}{\lambda_{\mathrm{cat}}}.
    \label{eqn:monopole_photon_yield}
\end{equation}
A monopole with $\hat{l} = 1$ will therefore produce the same number of Cherenkov photons per track length as a single-electric-charge, relativistic particle without stochastic energy losses along its track \cite{Radel:muonGEANT}.
This implies that $\hat{l}$ can be used to express the resulting monopole flux limits without assuming a specific decay channel (Sec. \ref{sec:flux_limits}).
This ansatz is valid as long as the monopole's light emission can be approximated as being continuous.
This condition is satisfied for a mean free path much smaller than the detector spacing.
From an experimental point of view  the speed $\beta$ and the mean free path $\lambda_{\mathrm{cat}}$ are the characterizing parameters for the detection  of such  monopoles.

Searches for slow monopoles based on the Rubakov-Callan effect have been pioneered with the underground detectors IMB and Kamiokande-II \cite{BeckerSzendyIMB,FukugitaKamiokandeII}
and the underwater detectors in Lake Baikal \cite{DomogatskyGyrlanda-86,BezrukovNT36,BelolaptikovNT36}.
A similar search has also been performed with AMANDA, the predecessor of IceCube \cite{PohlAMANDA}.

During the commissioning of the full detector (IC-86) in May 2011, a dedicated
trigger for slow particle signatures (Slow-Particle-Trigger, Sec. \ref{sec:slop}) in DeepCore was implemented. 
The denser instrumentation of DeepCore allows IceCube to detect monopoles of low light emission,
i.e. with rather large values of mean distances $\lambda_{\mathrm{cat}}$ between induced catalysis points.
In 2009, the deployment of the first DeepCore strings was still ahead.
Due to the larger spacing and the lack of an appropriate trigger,
IC-59  was blind for large $\lambda_{\mathrm{cat}}$.
For smaller $\lambda_{\mathrm{cat}}$ the mentioned drawbacks
were balanced by the larger geometrical area compared to DeepCore.

\subsection{Simulation of Magnetic Monopoles}\label{sec:simulation_monopole}
The signal expectation was determined from Monte Carlo simulations of magnetic monopoles in IceCube,
while the background expectation was determined from experimental data itself, with only supplementary simulations.

IceCube simulation includes particle injection and propagation, taking into account appropriate particle interactions, as well as the full detector response to the generated Cherenkov photons.

The arrival directions of magnetic monopoles are assumed to be isotropic.
The starting points of simulated monopole tracks are generated randomly on a disc of fixed size.
The distance of the plane is fixed with respect to the DeepCore detector but its orientation is random.
It is assumed  that the magnetic monopoles are not substantially
decelerated along their track
and  their velocity is constant \cite{Derkaoui:MonopoleIsotropy}.

The distances between the catalyzed nucleon decays 
 are simulated as a Poisson process with 
a mean free path $\lambda_{\mathrm{cat}}$ along the monopole track.
 Each nucleon decay is simulated as an electromagnetic cascade with an energy of $940\,\mathrm{MeV}$, corresponding to the benchmark detection channel (Eq. \ref{eqn:promising_decay_channel}). 
The simulation and propagation of the Cherenkov light from these  
cascades is done with the software package \textit{Photonics} \cite{Lundberg:Photonics} using the ice model described in \cite{IceCube:IceModel_AHA} for the IC-59 analysis and an improved version described in \cite{IceCube:SPICEMie} for the IC-86/DeepCore analysis.

Background noise in the DOMs has to be superimposed on the signal. This noise consists of uncorrelated  random noise, mostly from radioactive decays 
in the DOMs and correlated noise because of after pulses and signals from 
atmospheric muons. For the IC-59 analysis, the random 
noise is simulated as a Poisson process and the atmospheric muons are simulated using the software package CORSIKA \cite{CORSIKA} based on a 5-component model for cosmic rays with the hadronic interaction model SIBYLL \cite{Stanev:SIBYLL} and the H\"orandel flux model \cite{Hoerandel:Flux}.
For the simulation of noise in the IC-86/DeepCore analysis the detector response of simulated monopole signals is superimposed
with random and correlated noise hits from experimental data. These noise hits were recorded with a fixed rate trigger (FRT) that was implemented to measure and analyze background noise in the detector. More details on the FRT data are given in Sec. \ref{subsec:background_estimation}.

\begin{figure}
	\centering
	\includegraphics[width=0.9\columnwidth]{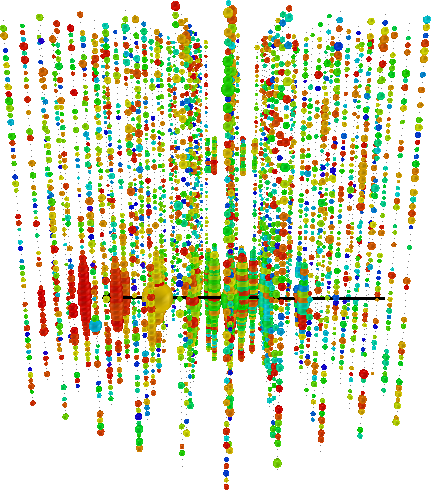}
	\caption{Event display of a simulated monopole with $\beta = 10^{-3}$ and $\lambda_{\mathrm{cat}} = 1\,\mathrm{cm}$ with superimposed background noise. The black line represents the monopole track. The DOMs are shown as tiny black dots. The color code illustrates the time scale from red for early times to blue for later times. The radii of the colored spheres scale with the number of recorded photoelectrons.}
	\label{fig:simulated_monopole}
\end{figure}

Figure \ref{fig:simulated_monopole} shows a simulated monopole event with $\beta = 10^{-3}$ and $\lambda_{\mathrm{cat}} = 1\,\mathrm{cm}$.
Because of the low speed,  the event duration for a monopole is
typically a factor of 1000 longer than for muon events and
 a large number of noise hits are recorded. However, the monopole also produces a large amount of Cherenkov light in the detector. 
Therefore, its signature can be separated from the randomly distributed noise hits already by eye.

\section{Search for Magnetic Monopoles with the Slow Particle Trigger}

The experimental data set was recorded between May 2011 and May 2012 with a dedicated slow particle 
trigger applied to DeepCore. In this period the livetime of the detector was 351 days, with a total number of approximately 50 million triggered events.

\subsection{The Slow-Particle-Trigger \label{sec:slop}}

Multiple IceCube triggers are implemented in the software of the
data acquisition system \cite{icecube:daq}. 
Most of them are sensitive to signatures of relativistic particles, e.g. muons,
so they have little sensitivity to non-relativistic magnetic monopoles.
Only for the case of very bright magnetic monopoles  
the large amount of light can frequently prompt triggers for relativistic particles. 
This case is described  in Sec. \ref{sec:ic59set}.

The Slow-Particle-Trigger (SLOP trigger) was first implemented in May 2011 \cite{Gluesenkamp:Monopoles}.
For the first year, the trigger operated only on the subdetector DeepCore.
Since May 2012, the trigger has been operating on the full IceCube detector.

\begin{figure}
	\centering
	\includegraphics[width=0.9\columnwidth]{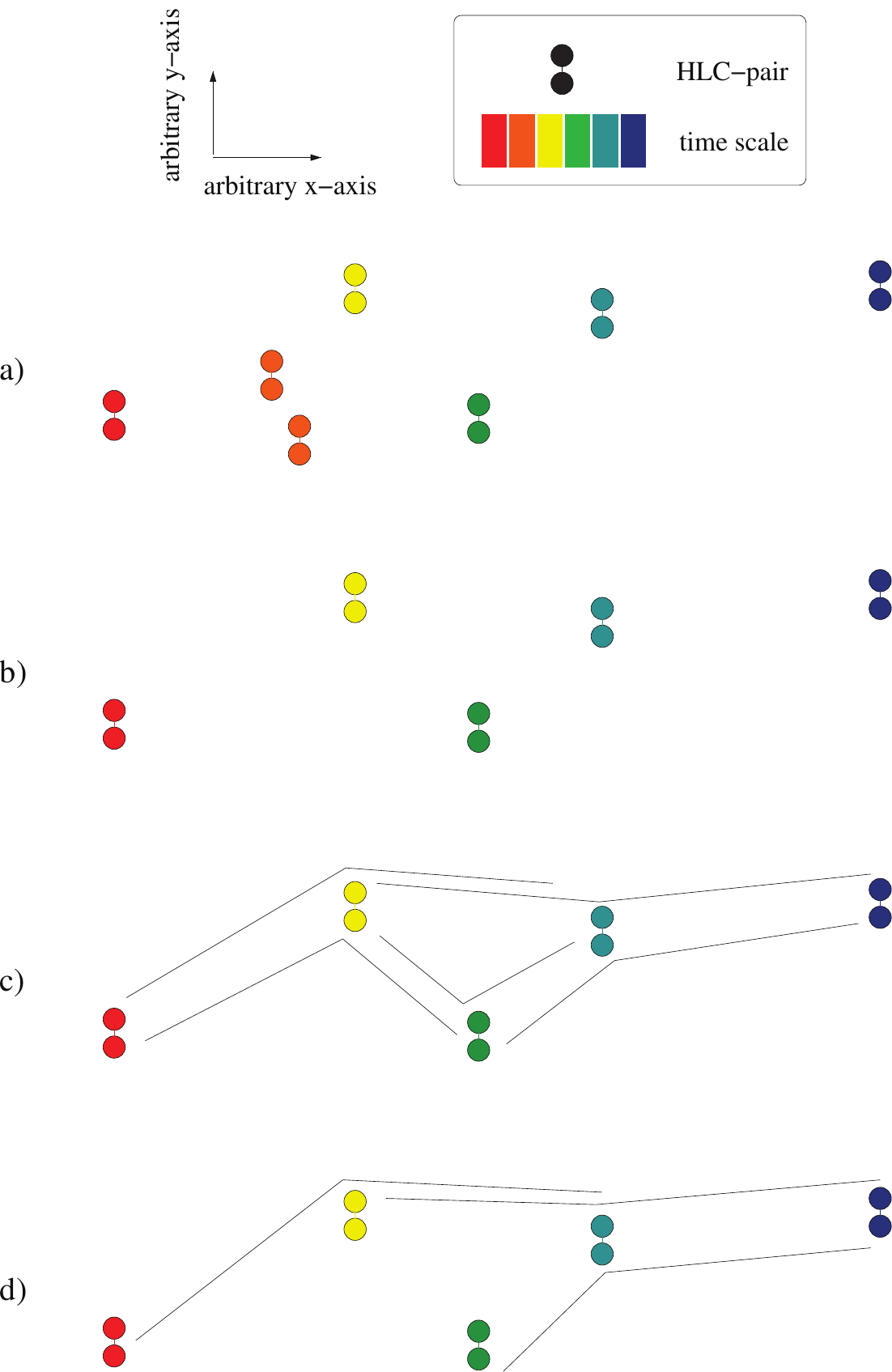}
	\caption{Illustration of the SLOP trigger. The times and positions are arbitrary.
             The x- and y-axis correspond to spatial coordinates and the color bar corresponds to a time scale.
             a) List of all HLC pairs. For the trigger algorithm only the position and time of the first hit of each HLC pair is used.
             b) The two HLC pairs (orange) with a time difference $\Delta t < t_{\mathrm{proximity}}$ are removed.
             c) All combinations of three HLC pairs, called triplet, with a time difference $\Delta t_{\mathrm{ij}} \in [t_{\mathrm{min}},t_{\mathrm{max}}]$ between two pairs are built.
             d) The cuts on the quality criteria $\Delta d$ and $v_{\mathrm{rel}}$ remove two more triplets.
             If the remaining triplets overlap in time and fulfill $n\mathrm{-triplet} \geq n_{\mathrm{min}}\mathrm{-triplet}$, a trigger is generated and the full detector data within the time span 
             from the first to the last HLC pair of the triplets is recorded. \cite{Gluesenkamp:Monopoles}}
	\label{fig:triplet_selection}
\end{figure}

The SLOP-Trigger searches for time isolated local 
coincidences in nearby DOMs caused by subsequent nucleon decays along the monopole trajetory.
These coincidences have to be consistent with a straight 
particle track of constant speed. 

The SLOP-Trigger is illustrated in Fig. \ref{fig:triplet_selection}. Specific values for the different trigger parameters are listed in Table \ref{tab:trigger_conditions}.
It is based on local coincidences of hits (HLCs, Sec. \ref{icecube}).
For the trigger, the position and time, defined by the first hit of the 
HLC pair, of all HLC pairs are stored in a list (Fig. \ref{fig:triplet_selection}a).
Since muons pass the detector within $\sim 5\,\mathrm{\mu s}$ they produce several HLC pairs within a short time.
By removing all HLC pairs with time differences $\Delta t < t_{\mathrm{proximity}}$ from the list, muon hits are efficiently rejected (Fig. \ref{fig:triplet_selection}b).

\begin{figure}
	\centering
	\includegraphics[width=0.9\columnwidth]{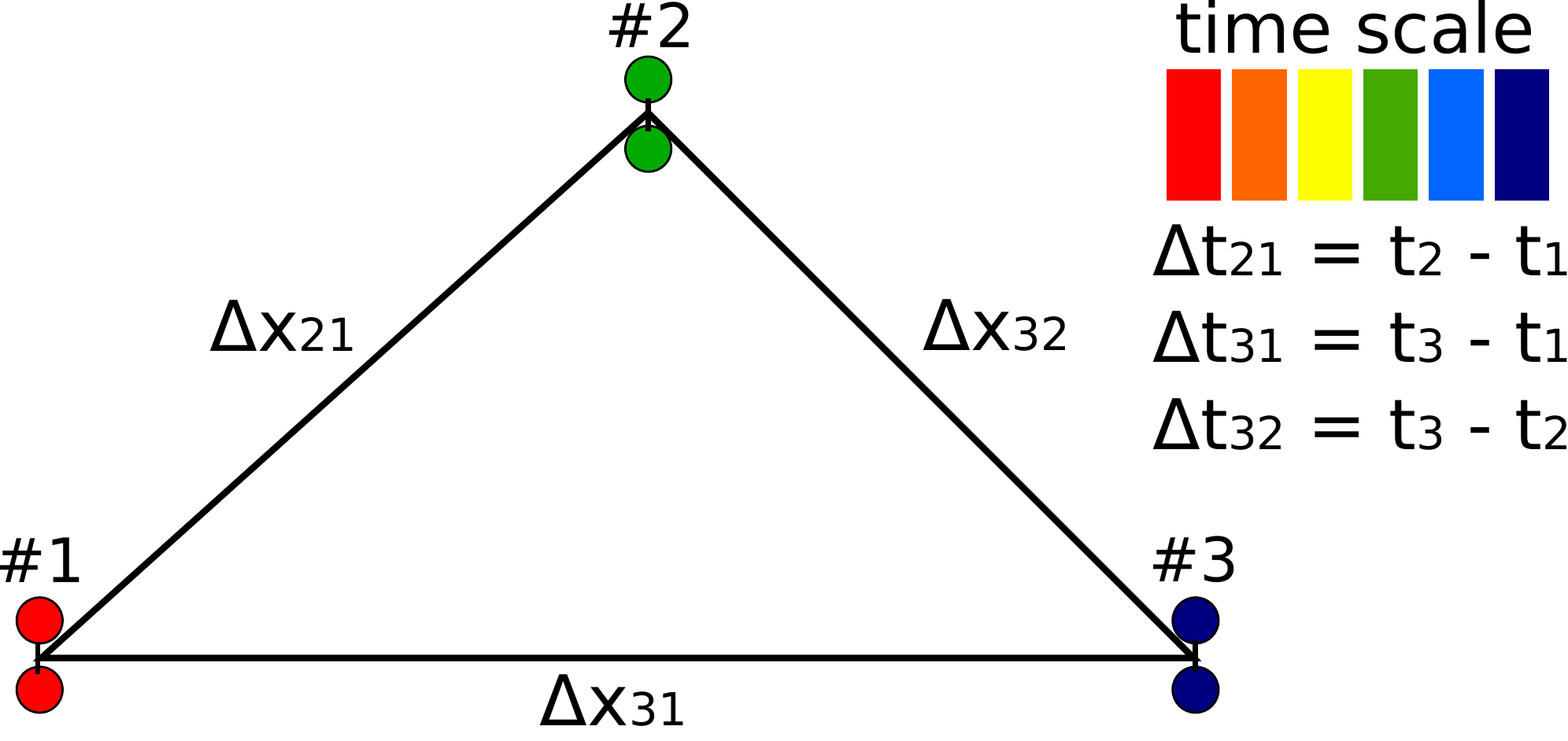}
	\caption{Illustration of a triplet. All three HLC pairs are defined by the position $(\vec{x_1},\vec{x_2},\vec{x_3})$ and the time $(t_1,t_2,t_3)$ of the first hit of an HLC pair. The trigger observables are  
the distances $(\Delta x_{21},\Delta x_{32},\Delta x_{31})$ and time differences $(\Delta t_{21},\Delta t_{32},\Delta t_{31})$.}
	\label{fig:triplet}
\end{figure}

The remaining HLC pairs are searched for every combination of three HLC pairs, the triplets (Fig. \ref{fig:triplet_selection}c).
 The time difference between any two HLC pairs 
within a triplet has to be in the range $[t_{\mathrm{min}},t_{\mathrm{max}}]$.
Furthermore only triplets that match a track-like signature are kept. Therefore two quality criteria are required:
the contributing HLC pairs have to be ordered along a line and the 
time differences  have to be consistent with a constant speed (Fig. \ref{fig:triplet}). 

The first can be verified by the parameter $\Delta d = \Delta x_{21} + \Delta x_{32} - \Delta x_{31}$. If $\Delta d = 0$ all HLC pairs are located on a line. The 
second can be checked by the parameter

\begin{equation}
	v_{\mathrm{rel}} = \frac{\left| \frac{1}{v_{21}} - \frac{1}{v_{32}} \right|}{\frac{1}{3} \cdot \left( \frac{1}{v_{21}} + \frac{1}{v_{32}} + \frac{1}{v_{31}} \right)},
\end{equation}
where $v_{ij} = \frac{\Delta x_{ij}}{\Delta t_{ij}}$ with $i,j\in\{1,2,3\}$ corresponds to the speed between the j-th and the i-th HLC pair within a triplet. For a monopole with a constant speed all HLC pairs should be connected by a constant speed and therefore $v_{\mathrm{rel}} \rightarrow 0$ should be valid. 
All triplets not satisfying these quality criteria are removed from the set of triplets (Fig. \ref{fig:triplet_selection}d).

Finally, if the number of triplets in the set overlapping in time, $n\mathrm{-triplet}$, is greater than a minimum number of triplets $n_{\mathrm{min}}\mathrm{-triplet}$, the trigger is launched.
When these conditions are met, the full detector data
 from the first to the last HLC pair in the list of  triplets are stored,
also including those DOM signals not contributing to the trigger.
 The maximum event duration of the trigger is restricted to 
$L_{\mathrm{max}} =5\,\mathrm{ms}$.

\begin{table}
	\centering
	\caption{Trigger conditions of the SLOP-Trigger \cite{Gluesenkamp:Monopoles}.}
	\label{tab:trigger_conditions}
	\begin{tabular}{ll}
		\hline\noalign{\smallskip}
		parameter & value \\
		\noalign{\smallskip}\hline\noalign{\smallskip}
		$t_{\mathrm{proximity}}$ & $ 2.5\,\mathrm{\mu s}$ \\
		$t_{\mathrm{min}}$ & $0\,\mathrm{\mu s}$ \\
		$t_{\mathrm{max}}$ & $500\,\mathrm{\mu s}$ \\
		$\Delta d$ & $\le 100\,\mathrm{m}$ \\
		$v_{\mathrm{rel}}$ & $\le 0.5$ \\
		$n_{\mathrm{min}}\mathrm{-triplet}$ & $ 3$ \\
		$L_{\mathrm{max}}$ & $5\,\mathrm{ms}$ \\
		\noalign{\smallskip}\hline
	\end{tabular}
\end{table}

\subsection{Background Study for the SLOP Data}\label{subsec:background_estimation}
To investigate the characteristics of the SLOP events, we use an experimental data set of $\sim 2$ days of live-time.
This is sufficiently short to exclude a significant signal contamination given by current flux limits (Sec. \ref{sec:intro})
and hence the data can be considered as background.

\begin{figure}
	\centering
	\includegraphics[width=0.9\columnwidth]{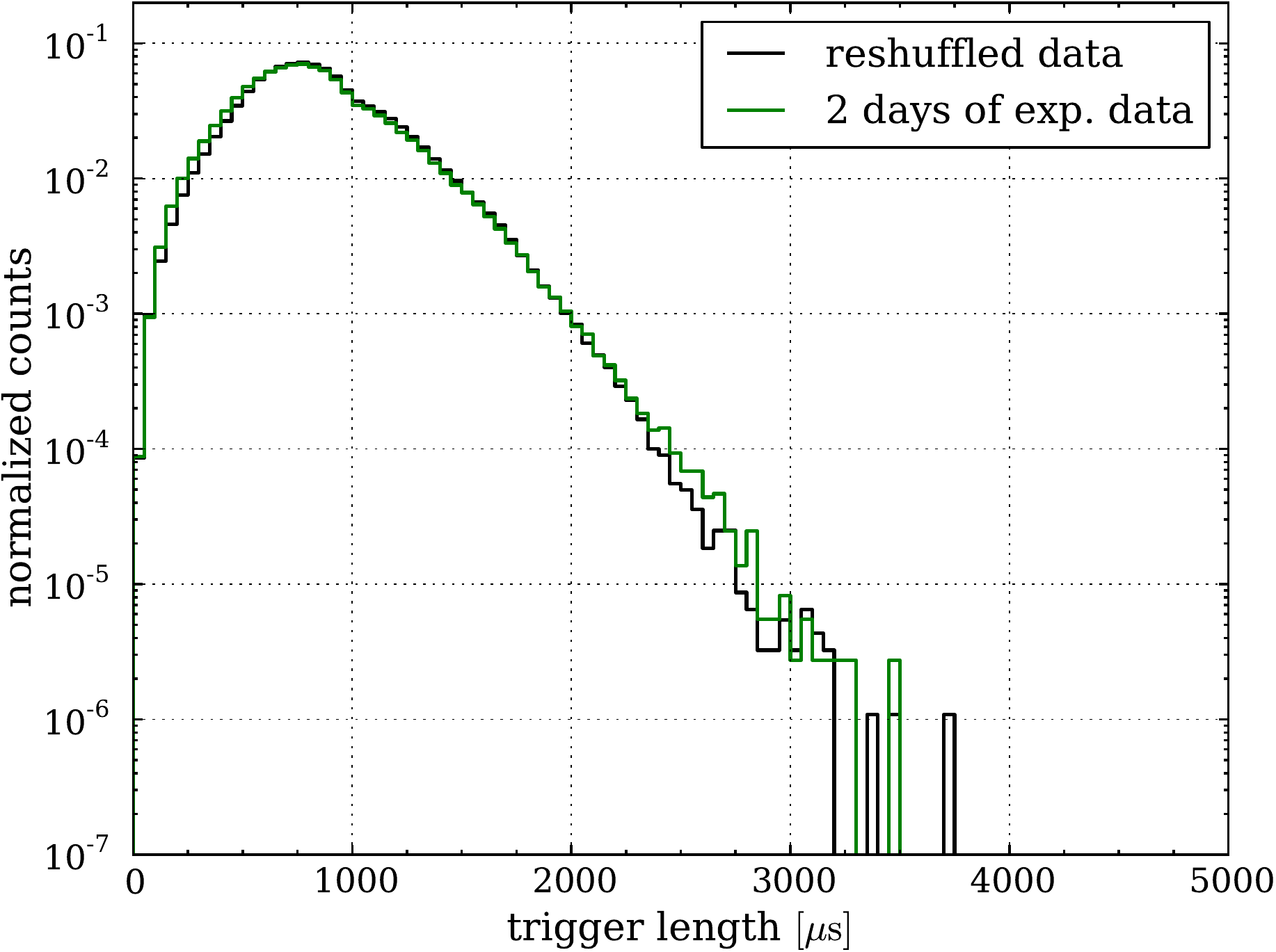}
	\caption{Event duration distribution of a experimental two days data 
set (green). The trigger rate is $2.1\,\mathrm{Hz}$. The maximum is at about $750\,\mathrm{\mu s}$. For comparison the event duration distributions 
of the generated background events (black) is superimposed. The number of entries is normalized to one.}
	\label{fig:trigger_length}
\end{figure}

Figure \ref{fig:trigger_length} shows the distribution of event durations
of SLOP triggered events. Typical durations are of the order of milliseconds, 
whereas other IceCube triggers have typical durations of a few microseconds. 

\begin{figure}
	\centering
	\includegraphics[width=0.9\columnwidth]{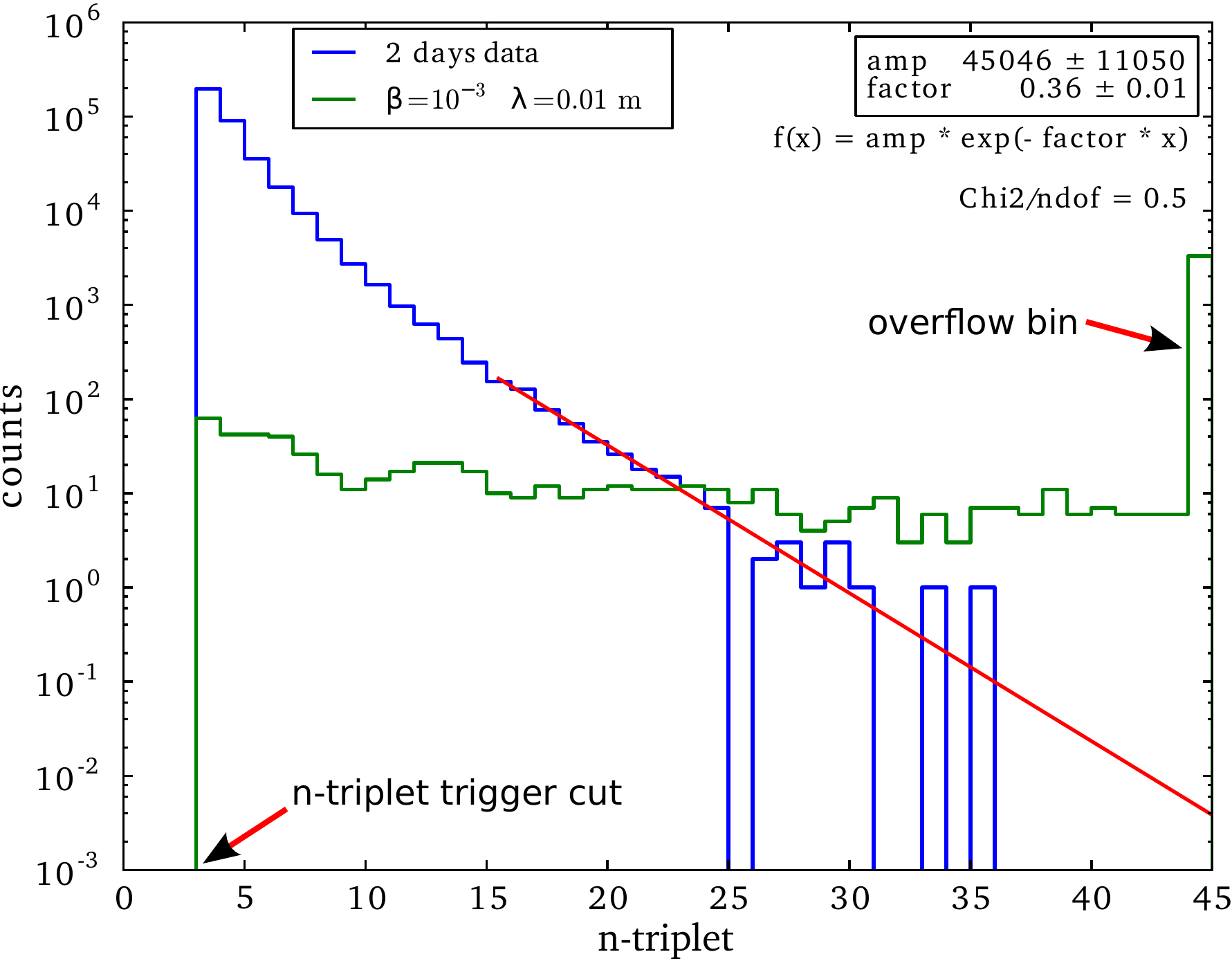}
	\caption{$n\mathrm{-triplet}$ distribution of the experimental test data sample (blue) in comparison to a distribution of simulated monopoles with 
             $\beta = 10^{-3}$ and $\lambda_{\mathrm{cat}} = 1\,\mathrm{cm}$ (green).
             In addition, an exponential function is fitted to the tail of the experimental distribution for $n\mathrm{-triplet} \geq 15$ (red).}
	\label{fig:2days_ntriplet}
\end{figure}

Figure \ref{fig:2days_ntriplet} compares the $n\mathrm{-triplet}$ distribution 
of the experimental data sample with simulated monopoles of $\beta = 10^{-3}$ and $\lambda_{\mathrm{cat}} = 1\,\mathrm{cm}$.
While the background distribution decreases rapidly for larger $n\mathrm{-triplet}$, the signal distribution is almost flat. 
Therefore, the quantity $n\mathrm{-triplet}$ discriminates well
 between signal and background events.
The exponential decrease of the background distribution indicates
a possible  Poissonian random process for 
combinations of HLC pairs which result in a triplet.

To understand the underlying random processes for background events
we developed a method to generate a high statistics sample of 
background events by reshuffling experimental events recorded with the FRT.
The FRT fires at fixed time intervals (e.g., every thirty seconds), and DOM data from the entire detector are recorded over a time interval of 10ms.
The resulting events contain all types of random and correlated backgrounds, and highly unlikely any signal.

\begin{figure}
	\centering
	\includegraphics[width=0.9\columnwidth]{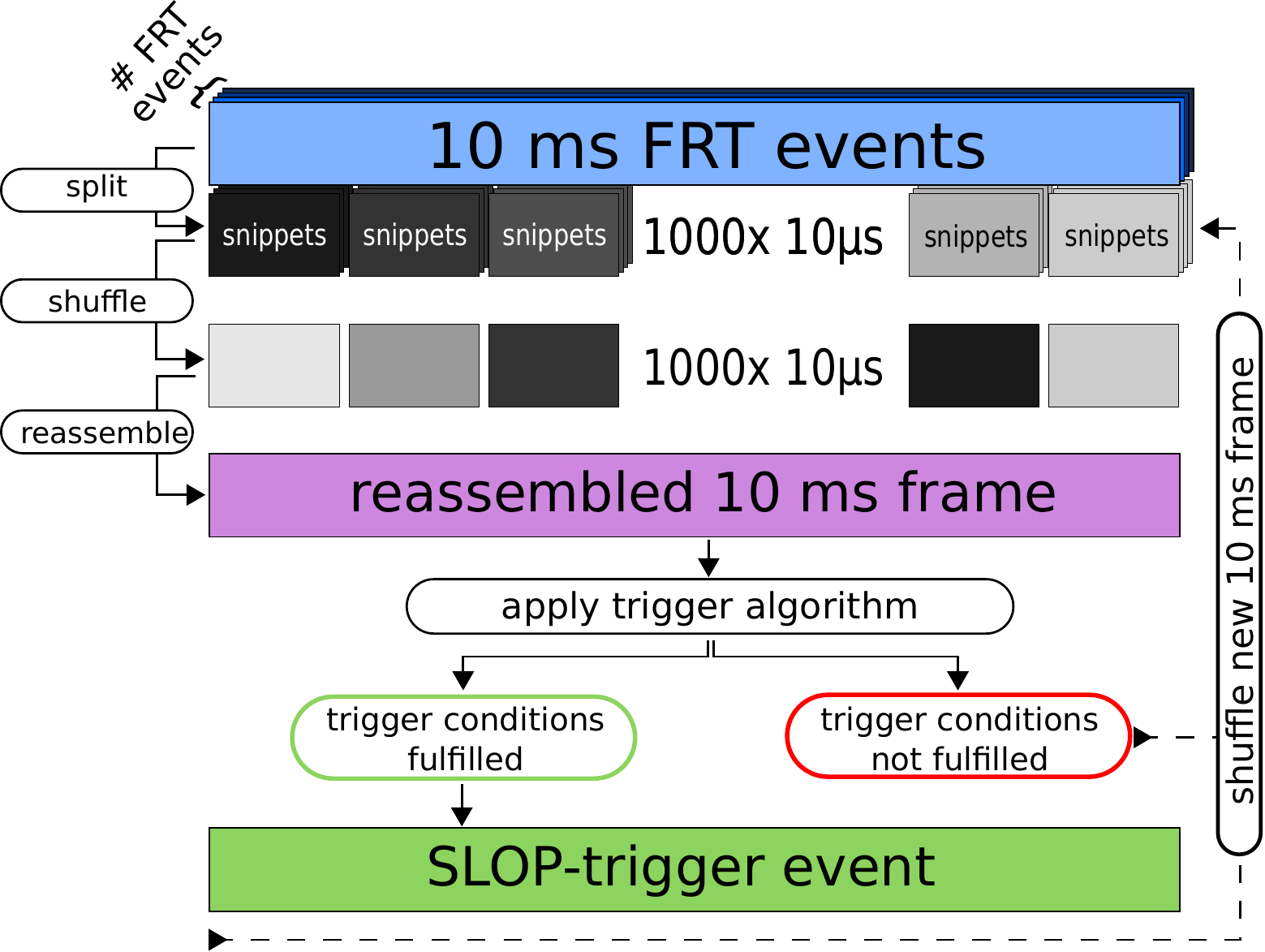}
	\caption{Illustration of the generation of background events by reshuffling  experimental data measured by a fixed rate trigger (FRT). FRT events have a length of $10\,\mathrm{ms}$. They are split into $10\,\mathrm{\mu s}$ snippets. The snippets are shuffled  randomly to build new $10\,\mathrm{ms}$ frames. Then the SLOP trigger algorithm is applied.}
	\label{fig:reshuffling_method}
\end{figure}

The FRT events of $10\,\mathrm{ms}$ length were split into snippets of $10\,\mathrm{\mu s}$, which were then randomly re-ordered to form new $10\,\mathrm{ms}$ events.
The newly assembled events are then passed to the SLOP trigger algorithm (Fig. \ref{fig:reshuffling_method}).
This way, a total of 400 seconds of FRT data were re-shuffled to generate a background sample of about 25 days of live-time equivalent.
The generated sample closely resembles the experimental SLOP-triggered events.
Figure \ref{fig:trigger_length} compares the event duration of the generated background events to the SLOP-triggered events in two days of experimental data.
The method reproduces the measured event duration distribution resonably well over  several 
orders of magnitude. For shorter event duration the distribution of the generated data sample tends to be below the distribution of the experimental test data sample.
This is expected because this method cannot correctly model noise hits that are correlated over time scales larger than the length of the $10\,\mathrm{\mu s}$ snippets.
Below $10\,\mathrm{\mu s}$ the triplets are characterized by the same DOM combinations due to the low statistics of the FRT events.
The overall good agreement indicates that correlated noise is a subdominant effect and  is only relevant for short time-scales.
We will presume later that different triplets due to correlated noise are themselves based on largely independent sets of HLC pulses.
Therefore, for large values of $n\mathrm{-triplet}$ the contribution from correlated noise triplets is added as a random process similar to the triplets from uncorrelated noise.

\begin{figure}
	\centering
	\includegraphics[width=0.9\columnwidth]{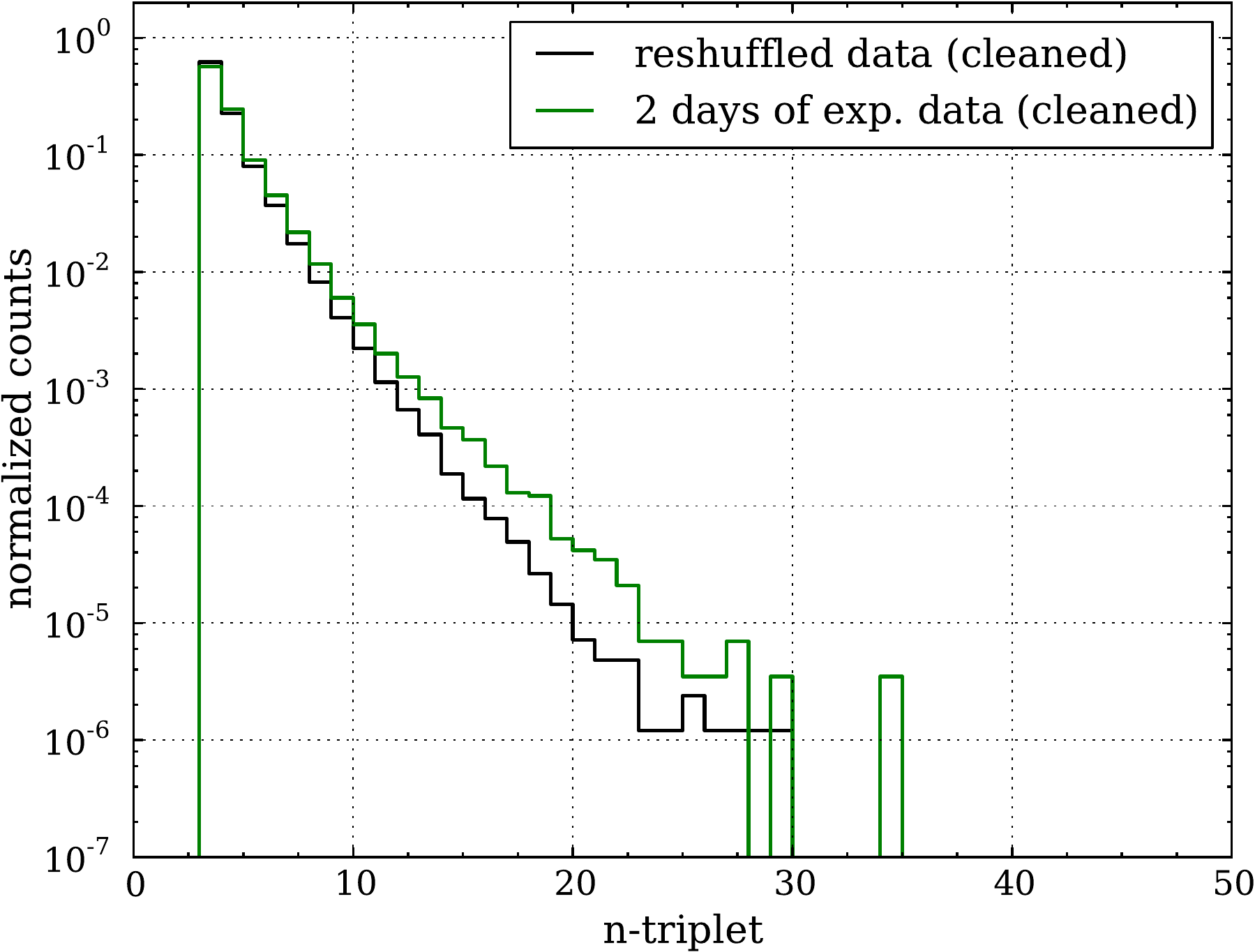}
	\caption{Comparison of the $n\mathrm{-triplet}$ distributions of the experimental test data set (green) and the generated background events (black). Triplets caused by HLC pairs fulfulling $\Delta t_{21}$ or $\Delta t_{32} \leq 50\,\mathrm{\mu s}$ are not taken into account (cleaned).}
	\label{fig:ntriplet_cleaned_frt}
\end{figure}

Figure \ref{fig:ntriplet_cleaned_frt} compares the $n\mathrm{-triplet}$ distributions of experimental data and generated background.
Overall both distributions are similar and show an exponential decay.
The differences can be understood by two effects. The first is the aforementioned effect that noise correlations over time scales longer than $10\,\mathrm{\mu s}$ are not taken into account,
which is expected to increase the number of triplets.
By removing triplets which arise from HLC-pairs fulfilling the typical time-scale of correlated noise ($\Delta t_{21}$ or $\Delta t_{32} \leq 50\,\mathrm{\mu s}$) the agreement improves.
However overall correlated noise has only a small effect on these distributions.
More importantly, the FRT data and the SLOP test-data do not correspond to the same data taking period.
The DOM noise rate shows slow slight drifts over long periods of time.
The chance probability of producing large n-triplet values depends on this random noise.
This effect is accounted for by the background fit described in the following section. 
In conclusion, the observed background is understood by the noise characteristics of the DOMs.

\subsection{Background Model for the SLOP Data}\label{sec:backgroundmodel}

As result of the findings in the previous section, 
the $n$-triplet distribution for background is estimated by 
fitting the experimental data with a 
simple probabilistic model.

The generic ansatz assumes that the probability to find a triplet (3 HLC-pairs) can be  described with a combinatorial model.
For a number of $N$ HLC pairs the maximum number of possible 
triplets is given by 
$n_{\mathrm{max}}=\tbinom{N}{3}$. 
If the probability $p$ for any 
three out of $N$ HLC pairs to build a triplet is  approximately
 constant, then  the probability for $n$ triplets out of 
$n_{\mathrm{max}}$ possible triplets follows the binomial distribution:

\begin{equation}
	B\left(n \,\arrowvert\, n_{\mathrm{max}},p\right) = \binom{n_{\mathrm{max}}}{n} \, p^{n} \, (1-p)^{n_{\mathrm{max}}-n}.
\end{equation}
As the HLC-pairs themselves arrise from random noise, the
 probability to observe  $N$ HLC pairs is given by a 
Poissonian:

\begin{equation}
	P_{\mu}\left( N \right) = \frac{\mu^{N}}{N!} \, e^{- \mu},
\end{equation}
where $\mu$ is the mean expectation for  the number of HLC pairs $N$ 
 in a given time window. 
The total probability to observe the number $n$ triplets is given by a sum over all binomial probabilities $ B\left(n \,\arrowvert\, n_{\mathrm{max}}(N),p\right) $ weighted with the probability to observe $N$ HLC pairs.
This results in

\begin{equation}
	P(n \,\arrowvert\, \mu,p) = P_0 \sum\limits_{N = N_{\mathrm{min}}(n)}^{\infty} P_{\mu}\left( N \right) \cdot B\left(n \,\arrowvert\, n_{\mathrm{max}}(N),p\right),
\end{equation}
The sum starts with the minimum number of HLC-pairs  $N_{\mathrm{min}}(n)$ 
which are combinatorially required for $n $ triplets.
This is given by the solution of the equation 
$n = \binom{\lfloor N_{\mathrm{min}} \rfloor}{3}$.
 Here, $\lfloor N_{\mathrm{min}} \rfloor$ is the greatest integer less or 
equal to $N_{\mathrm{min}}$. 
The parameter $P_0$ is the total normalisation of 
$P(n \,\arrowvert\, \mu,p)$. 

\begin{figure}
	\centering
	\includegraphics[width=0.9\columnwidth]{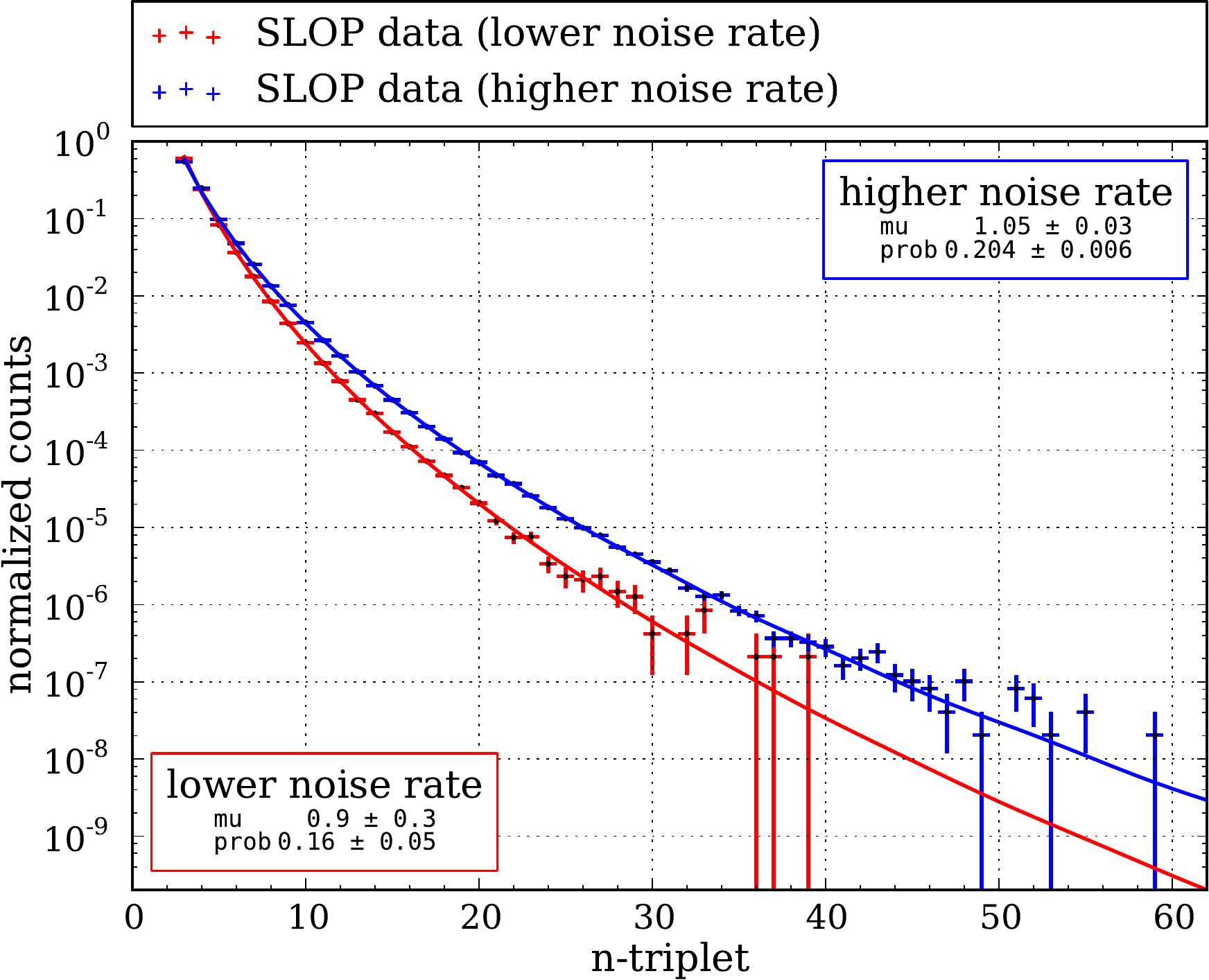}
	\caption{$n\mathrm{-triplet}$ distributions of experimental SLOP data. The blue distribution corresponds to a $\sim 15\%$ higher noise rate than the red one. Both distributions are normalized to one. The solid lines show the results of the background model fit and the fit parameters $\mu$ and $p$ are shown in the boxes.}
	\label{fig:ntriplet_fit}
\end{figure}

With this ansatz it is possible to describe the distribution of 
$n\mathrm{-triplet}$ with only three parameters $P_0$, $\mu$ and $p$.
Figure \ref{fig:ntriplet_fit} shows the fit of this model to two normalized, experimental $n\mathrm{-triplet}$ distributions
which are based on SLOP data corresponding to different noise rates.
Since the distributions are normalized only $\mu$ and $p$ have to be fit.
The background model well describes the $n\mathrm{-triplet}$ distributions over several orders of magnitude.
Moreover the increase in the noise rate is reflected in an increase of the value of $\mu$ which depends on the noise rate.
In summary it can be confirmed that the background events from the SLOP trigger are dominated by random noise.

\subsection{Reconstruction of a Monopole Track}

The analysis searches for monopoles from all directions. Also
the random background  is largely isotropic and the 
event selection does not depend specifically
on the direction of the monopole. However, an important observable is the 
speed of the  track, which can be estimated with the  \textit{linefit} \cite{Ahrens:AMANDA_Reconstruction}.
This algorithm is based on a simple ansatz
in which the geometry of the Cherenkov cone and
the  optical properties of the medium are ignored and the particle is assumed to
 travel
with a velocity $\mathbf{v}$ along a straight line through
the detector. 
A pseudo-$\chi^{2}$ is constructed with the positions $ \vec{x}_i $ 
and times $  t_i $ of the HLC pairs of all $N$ selected triplets:

\begin{equation}
	\chi^{2} = \sum\limits_i^N \frac{\left|\vec{x}_i - (\vec{x}_0 + \vec{v} \cdot t_i)\right|^2}{1\,\mathrm{m}^2}.
\end{equation}
HLC pairs which participate in multiple triplets are taken into account
 multiple times.
This $	\chi^{2}  $ can be minimized analytically with respect to the
 speed  $\vec{v}$ and vertex $\vec{x}_0 $. 
Note that  $ \chi^{2} $  is arbitrarily
normalized and cannot be interpreted statistically in terms of 
goodness of fit. The following analysis uses only the estimated speed $|\vec{v}|$.

\begin{figure}
	\centering
	\includegraphics[width=0.9\columnwidth]{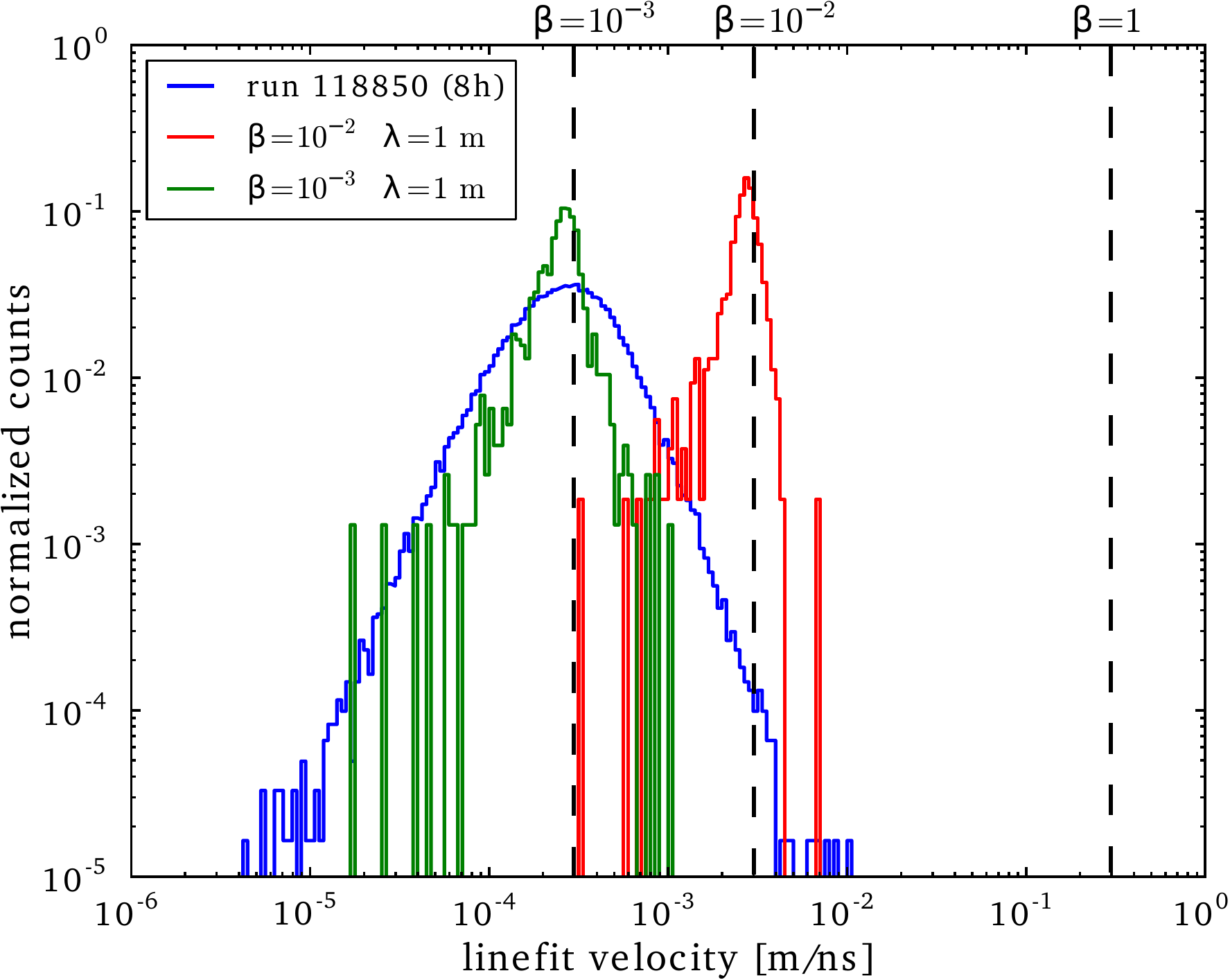}\\\vspace{0.25cm}
	\includegraphics[width=0.9\columnwidth]{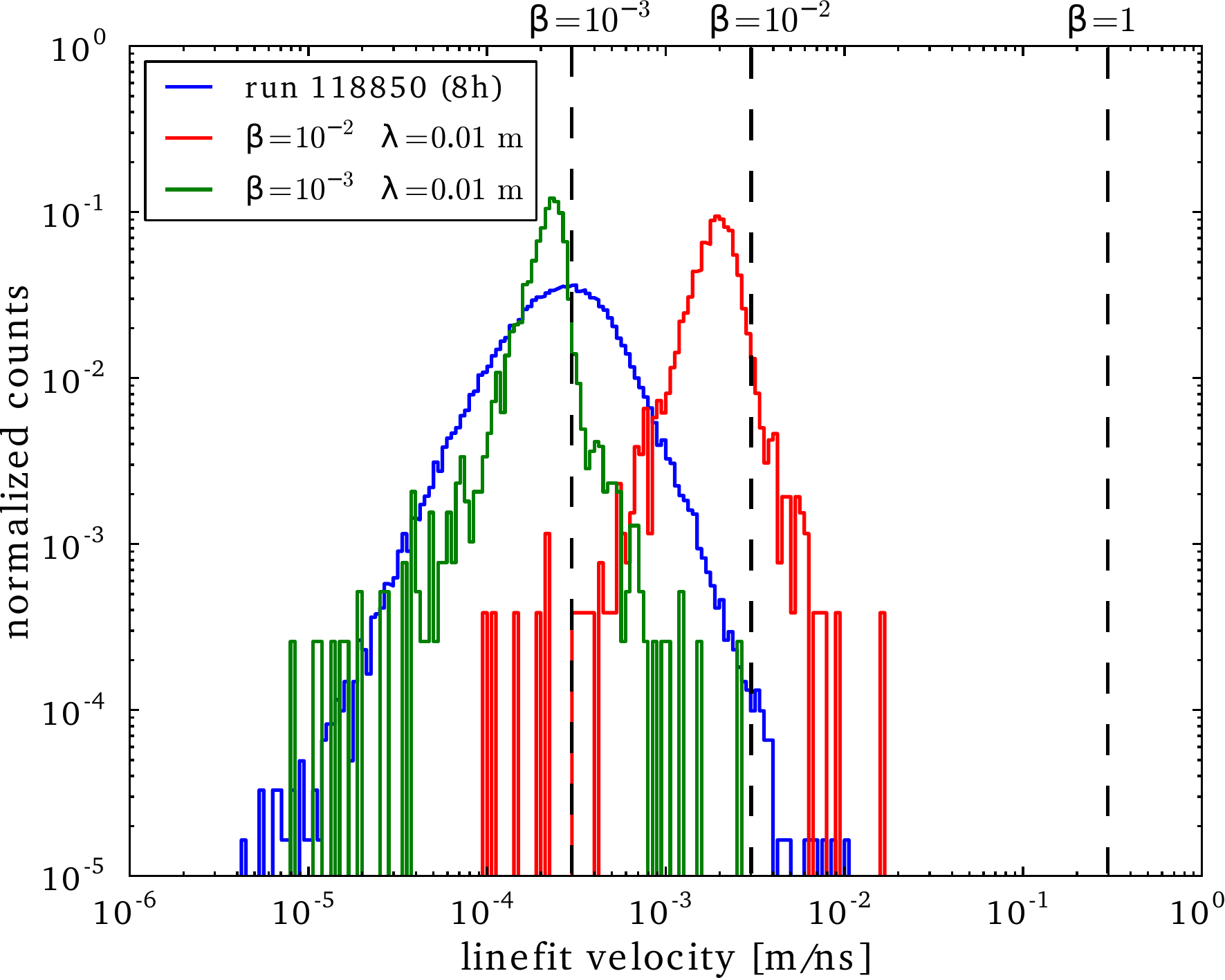}
	\caption{Distribution of the reconstructed speeds for two different simulated monopole speeds. At the top the distributions for monopoles with $\lambda_{\mathrm{cat}}=1\,\mathrm{m}$ and at the bottom for monopoles with 
$\lambda_{\mathrm{cat}}=1\,\mathrm{cm}$ are shown. For comparison, reconstructed
experimental  data corresponding to a live-time of 8h are plotted. 
The three dotted black lines show the true speeds and the speed of light. All distributions are normalized to one.}
	\label{fig:velocity_reco}
\end{figure}

In Fig. \ref{fig:velocity_reco} the distributions of the reconstructed speeds are shown. The reconstructed speeds are a reasonable 
estimate of the true speed, in particular for faint monopoles 
(see $\lambda_{\mathrm{cat}}=1\,\mathrm{m}$). 
For brighter monopoles (see $\lambda_{\mathrm{cat}}=1\,\mathrm{cm}$), the reconstructed speeds slightly  
underestimate the true speed. 

This reconstruction algorithm is simple, robust and fast, while still yielding a sufficient accuracy.
It also allows us to approximate the  monopole direction  by the direction of
 $\vec v$. The mean difference between the true and reconstructed direction
varies between $\sim 11^{\circ}$ and $\sim 20^{\circ}$ depending on the 
monopole speed and the mean free path $\lambda_{\mathrm{cat}}$.

\subsection{Event Selection and Background Reduction}\label{sec:SLOPeventselection}

For this first IceCube analysis of SLOP data a robust approach based on 
 $n\mathrm{-triplet}$ as single final selection criterion and the determination
 of the expected background from  experimental data was chosen.

\begin{figure}
	\centering
	\includegraphics[width=0.9\columnwidth]{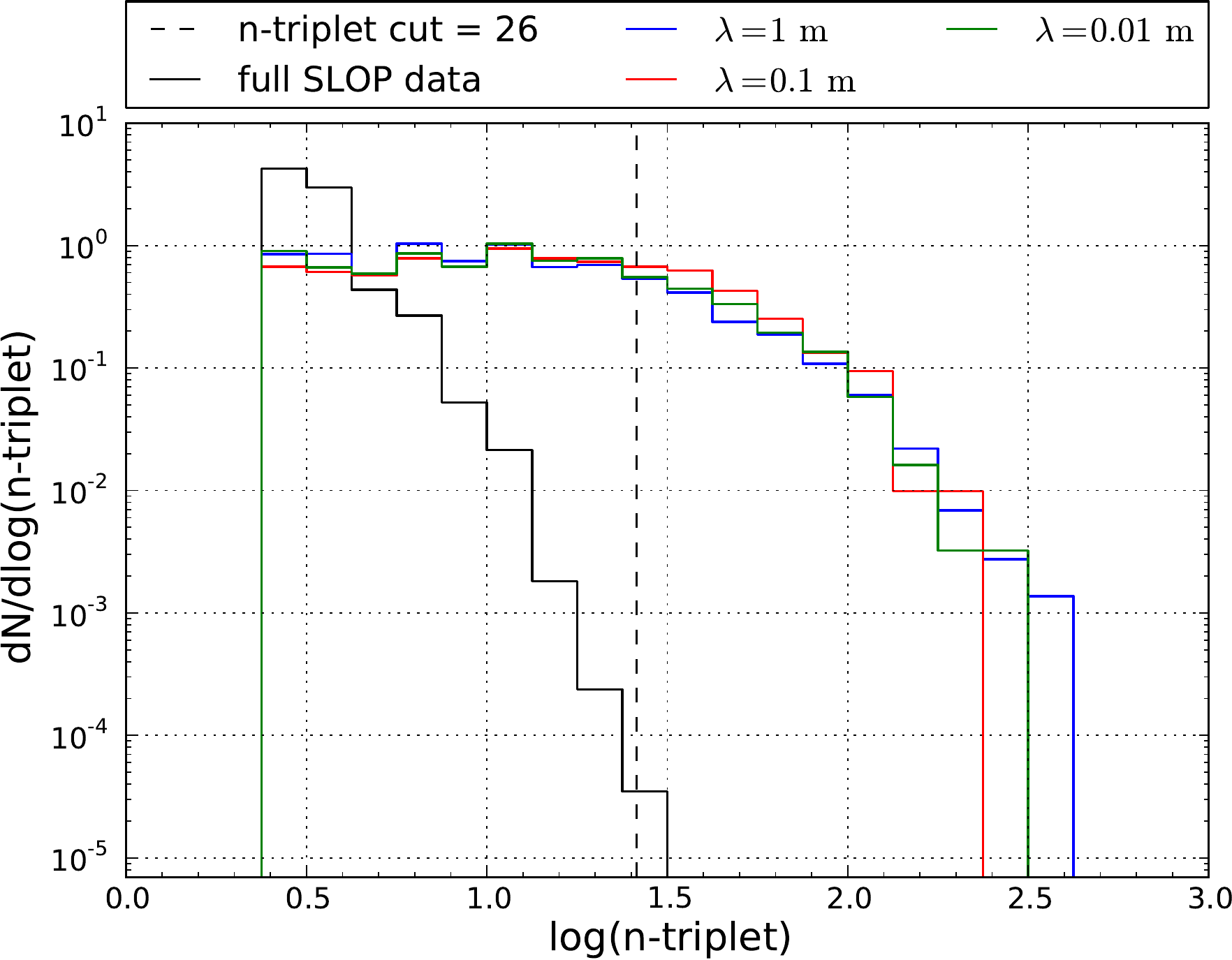}\\\vspace{0.25cm}
	\includegraphics[width=0.9\columnwidth]{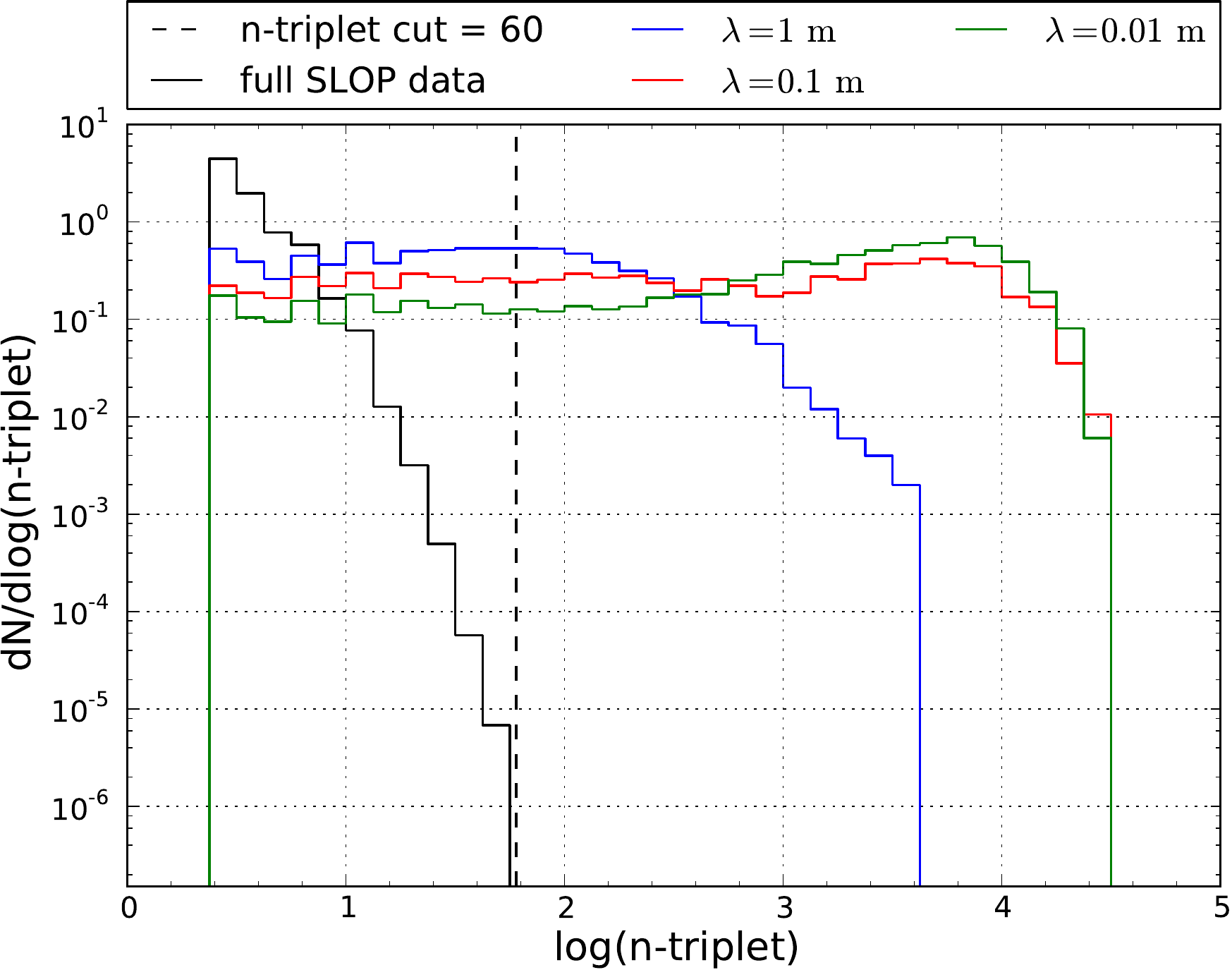}
	\caption{Probability density distributions of $n\mathrm{-triplet}$ for events with larger reconstructed speed (top) and for events with smaller reconstructed speed (bottom). In black the distributions of one year experimental data are shown. The signal distributions are shown with decreasing $\lambda_{\mathrm{cat}}$ in blue, red and green. The final cuts on $n\mathrm{-triplet}$ are shown by the dashed black line.}
	\label{fig:pdf_ntriplet_final}
\end{figure}

Figure \ref{fig:pdf_ntriplet_final} shows the probability density 
distributions of $n\mathrm{-triplet}$ for events with a 
reconstructed speed of at least 
$10^{-3}\,\mathrm{m/ns}$ (top) and with a reconstructed speed less than $10^{-3}\,\mathrm{m/ns}$ (bottom). 
While the signal expectation extends to very high $n\mathrm{-triplet}$, 
the distributions of the experimental data decrease rapidly. 
The final cuts on $n\mathrm{-triplet}$ were optimized for maximum sensitivity based on the  \textit{Model Rejection Factor} \cite{Hill:MRF}.
The optimization resulted in the following criteria/cuts:
$n\mathrm{-triplet}\ge 60$ for a reconstructed speed $v < 10^{-3}\,\mathrm{m/ns}$ and  $n\mathrm{-triplet}\ge 26$ for  $v \ge 10^{-3}\,\mathrm{m/ns}$.

These selection cuts were defined before unblinding the full experimental data.
Here, an iterative two step procedure was chosen. 
First, 10\% experimental data was unblinded with the selection 
determined by  the aforementioned  experimental 2 days data sample.
After no signal or unexpected background was observed the same 
procedure was applied to the full experimental data.

\begin{figure}
	\centering
	\includegraphics[width=1.0\columnwidth]{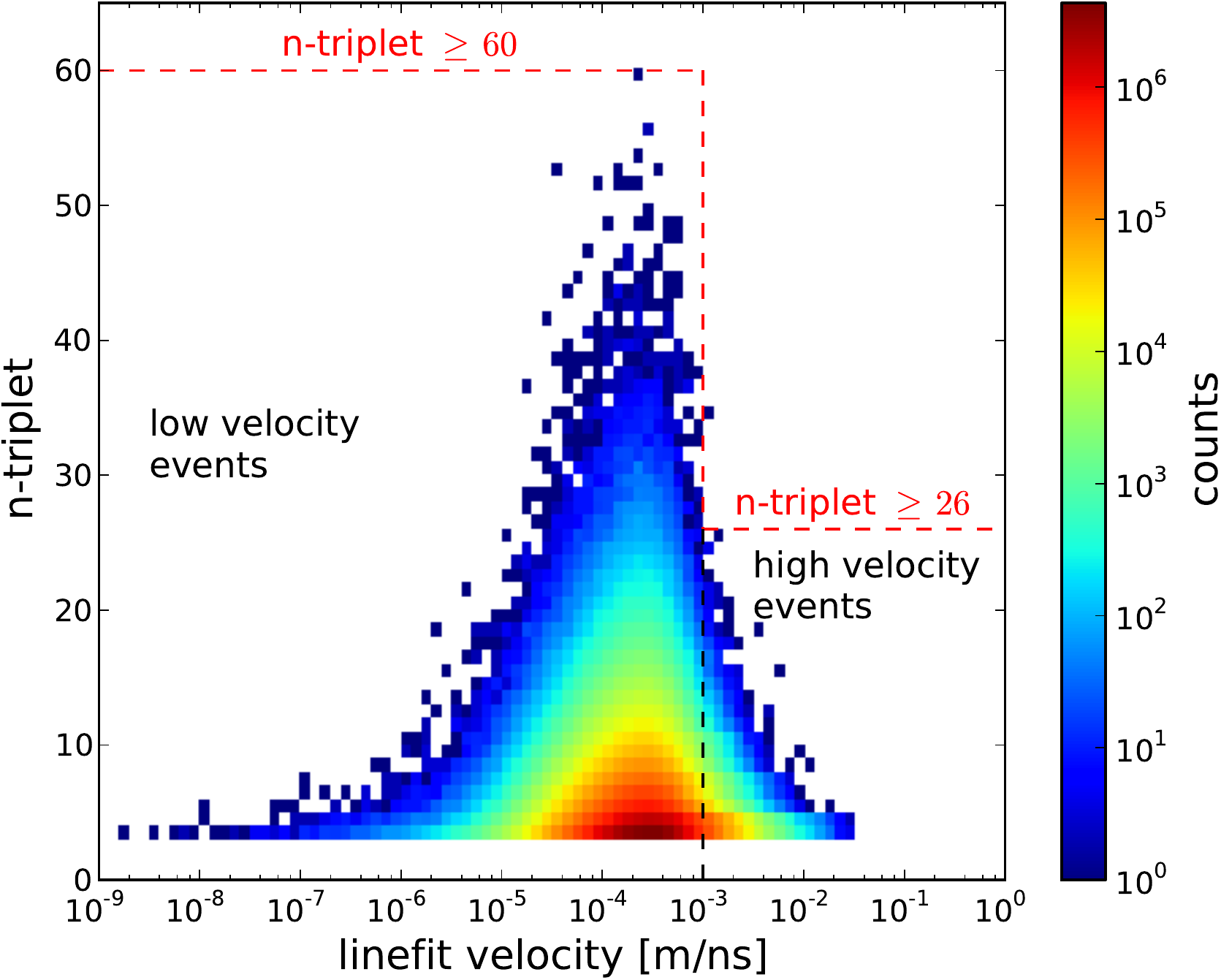}
	\caption{$n\mathrm{-triplet}$-speed distribution of one year of experimental data. The final cuts on $n\mathrm{-triplet}$ are shown by the dashed red lines. The boundary between the two speed regions is shown by the dashed black line.}
	\label{fig:velocity_ntriplet_final}
\end{figure}

Figure \ref{fig:velocity_ntriplet_final} shows the 
resulting $n\mathrm{-triplet}$-speed distribution 
for the full  year of experimental data. 
After the final selection only one experimental event with $n\mathrm{-triplet} = 34$ and $v = 1.15 \cdot 10^{-3}\,\mathrm{m/ns}$ remains, 
but not well separated from background. 
Closer inspection revealed no evidence for an obvious  
track-like signature,
in particular most triplets would not 
have survived tighter causality requirements. 
As this observation is consistent with the  expected number of about
$3$  background events (see below), we do not consider this result  
as positive detection.

\subsection{Results}\label{sec:slop_results}
\begin{figure}
	\centering
	\includegraphics[width=1.0\columnwidth]{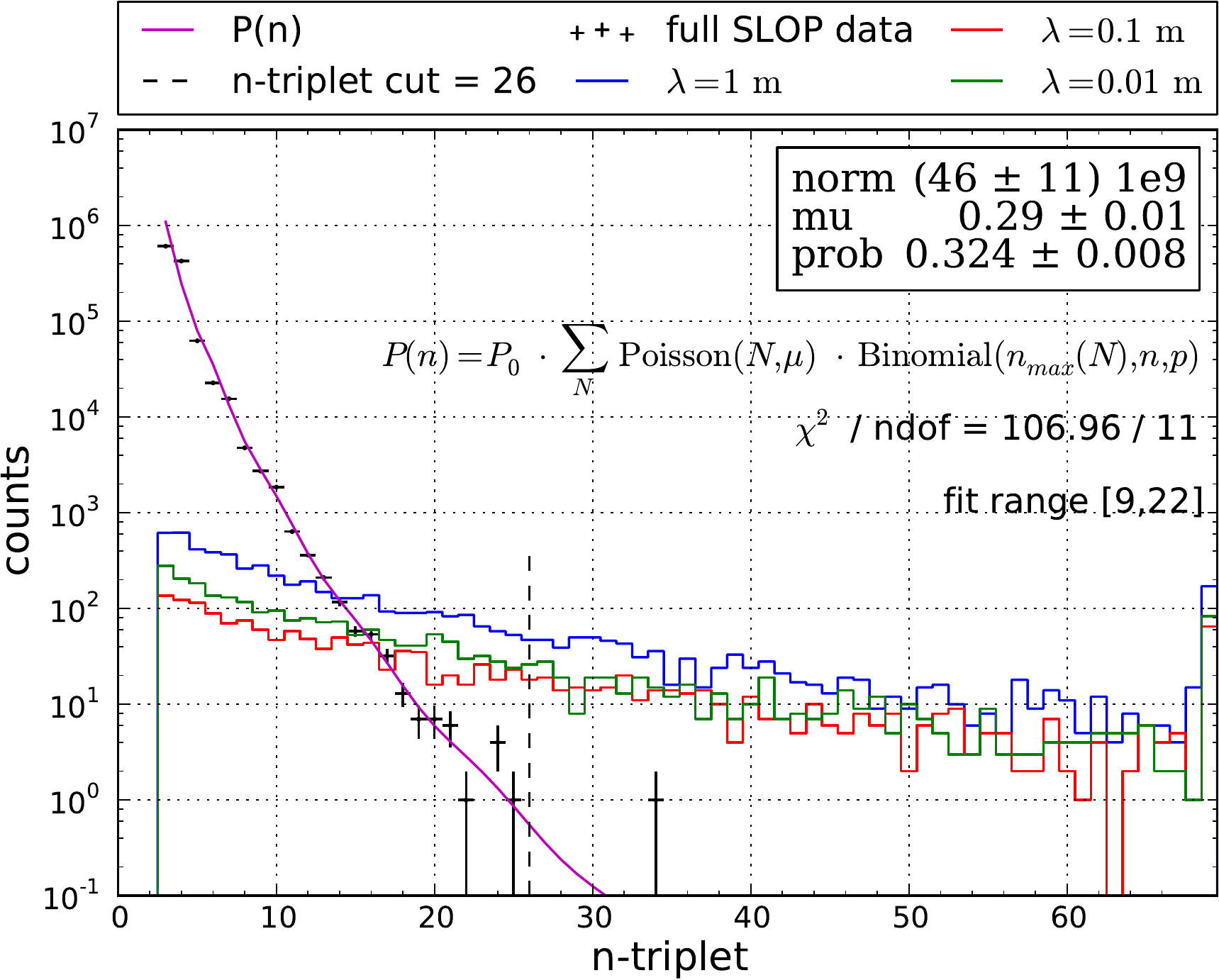}\\\vspace{0.25cm}
	\includegraphics[width=1.0\columnwidth]{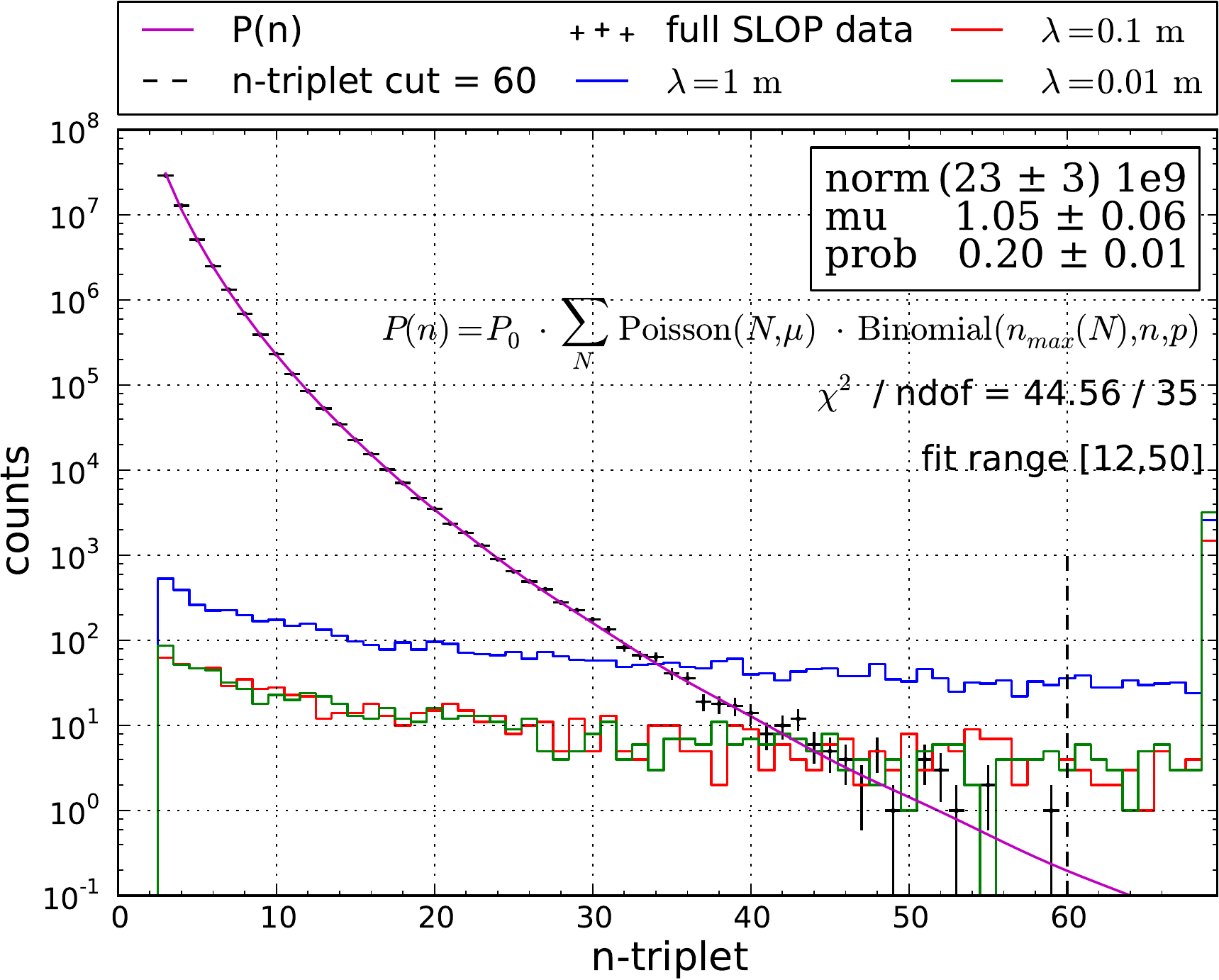}
	\caption{$n\mathrm{-triplet}$ distributions for events with larger reconstructed speed (top) and for events with a smaller reconstructed speed (bottom).
    The black data points show the distributions of the full experimental data.
    The expected signal is shown with decreasing $\lambda_{\mathrm{cat}}$ in blue, red and green.
    The fitted functions $P(n \,\arrowvert\, \mu,p)$ are shown in purple and the final selections on $n\mathrm{-triplet}$ as dashed black lines.}
	\label{fig:dist_ntriplet_final}
\end{figure}

With  no observed monopole signal we have  derived  an upper limit on 
the flux of  non-relativistic magnetic monopoles (Sec. \ref{sec:flux_limits}).
For this, the background model 
is fit to the $n\mathrm{-triplet}$ distributions for both speed ranges
(Fig. \ref{fig:dist_ntriplet_final}). It is found that the background model (Sec. \ref{sec:backgroundmodel}) well describes the $n\mathrm{-triplet}$ distribution over several orders of magnitude. 

\begin{figure}
	\centering
	\includegraphics[width=0.9\columnwidth]{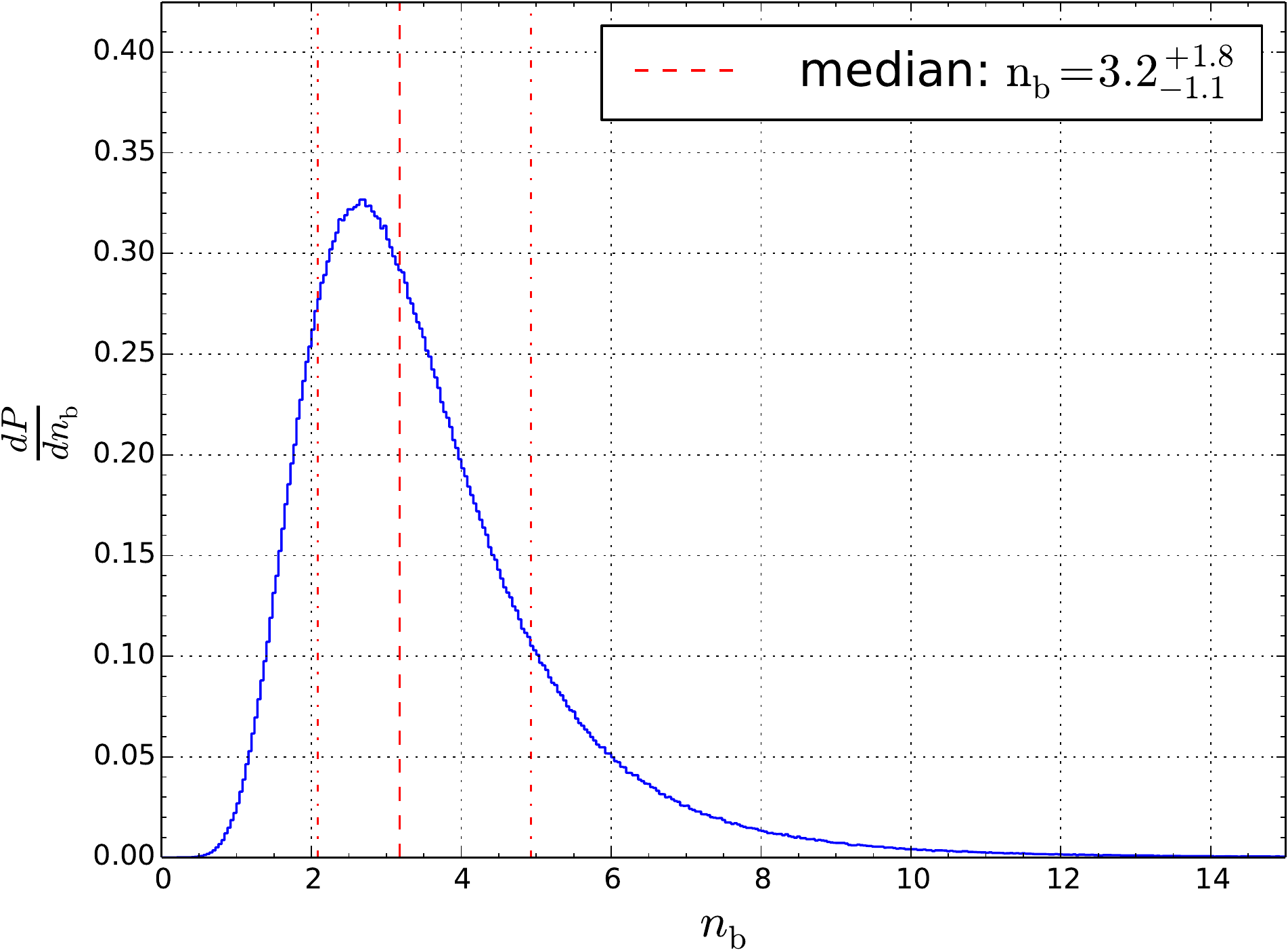}
	\caption{Probability density distribution of the expected number of background events. The median is shown by the dashed red line. The quantiles $Q_{0.16}$ and $Q_{0.84}$ are shown by the dash-dotted red line.}
	\label{fig:pdf_background}
\end{figure}

Based on the fit results, the expected number of background events with $n\mathrm{-triplet} \geq n_{\mathrm{cut}}$  can be calculated by the integral

\begin{equation}
	n^{\mathrm{fast/slow}}_{\mathrm{b}} = \int\limits_{n_{\mathrm{cut}}}^\infty P(n \,\arrowvert\, \mu^{\mathrm{fast/slow}},p^{\mathrm{fast/slow}}) \,dn.
\end{equation}
The total expected number of background events is defined by the sum of the expectation of both speed regions $n_{\mathrm{b}} = n^{\mathrm{fast}}_{\mathrm{b}} + n^{\mathrm{slow}}_{\mathrm{b}}$. By varying the fit-parameters within their fitted 
uncertainty a pseudo-experiment with different expected numbers of background events can be performed.
Figure \ref{fig:pdf_background} shows the resulting probability density distribution of the total expected number of background events.
The median expected number of background events is  $n^{\mathrm{median}}_{\mathrm{b}} = 3.2^{+1.8}_{-1.1}$. Here, the statistical uncertainty is approximated
 by the difference between the median and the quantiles $Q_{0.16}$ and $Q_{0.84}$.

\section{Search for Very Bright Magnetic Monopoles with the IC-59 Array}

The search for magnetic monopoles presented in this section uses data
taken during the season 2009-2010 when IceCube was running 
in its 59-string configuration. 
This analysis used the data taken with the standard IceCube triggers. 
The standard trigger that is used for highly energetic relativistic particles is a simple multiplicity trigger (SMT), which requires at least eight HLC hits within a sliding time window of 5 $\mu s$ (SMT-8). Other triggers are optimized for relativistic particles with lower energies. Data are recorded over at least the time interval over which the trigger condition of any of the triggers is fulfilled.  For HLC hits, the full PMT waveforms are digitized and recorded \cite{icecube:daq}.
Not all triggered events were transmitted to the Northern hemisphere by satellite.
Events of various categories (e.g. track-like, cascade-like, very bright events, etc.) 
have been selected by various online filters at the South Pole \cite{icecube:muonfilter,icecube:cascade_filter}.  
Although the filters are optimized for relativistic particles, they may accept bright monopole events if a sufficient number of DOMs are hit.
This analysis uses the cascade and high-energy filters, which have the best acceptance for non-relativistic monopoles.
The total livetime of this data set  is 311.25 days, with an average rate of
selected events of 85.5\,Hz.
The efficiency of this filter selection with respect to the multiplicity trigger
is above 75\% for monopoles of $\beta = 10^{-3} $ and
$\lambda_{\mathrm{cat}} = 1\,\mathrm{mm}$.

\subsection{Selection of Very Bright Magnetic Monopoles\label{sec:ic59set}}

Slow monopoles with a catalysis cross section $\sigma_{\mathrm{cat}}$ much
larger than $10^{-23}\,{\mathrm{cm^{2}}}$ appear as very bright tracks. 
Simulations of the detector response to such tracks show that the multiplicity condition is fulfilled over most of the monopole crossing time, or that successive triggers occur close enough in time for the recording intervals to overlap. So, a large fraction of a monopole's catalysis signature would be captured in a single event, if $\sigma_{\mathrm{cat}}$ is sufficiently high.
For  $\sigma_{\mathrm{cat}} < 10^{-23}$ cm$^{2}$ monopoles still yield multiple triggers, but the triggers occur less frequently, so that the signature is often split up into several sub-events.
The smaller the cross section, the more the monopole event
splits up and the larger are the gaps between the sub-events. 
Eventually, the signal becomes indistinguishable from the background. 
Therefore, this analysis 
focuses on catalysis cross sections above $10^{-23}$ cm$^{2}$.
For monopoles with such high  $\sigma_{\mathrm{cat}}$, the IC-59 analysis achieves a better sensitivity than the analysis using the SLOP trigger. This is simply because the IC-59 array had a much larger detection volume than the DeepCore array available to the previously described analysis. Future monopole searches will use data taken after 2012, when the SLOP trigger was operating on the full IC-86 array. These analyses will take advantage of both the large detection volume of the full IC-86 array and the high efficiency of the SLOP trigger.

\subsection{IC-59 Background Reduction}

The high-energy and cascade filters provide a data sample with about $10^9$ events.
The vast majority of these events are down-going atmospheric muons. This background is reduced using 
a set of straight cuts in a first step. These cuts are based on the time and location of the detected Cherenkov photons. Contrary to the IC-86/DeepCore analysis, whose cut parameters where defined using the time and location of DOM launches or HLC pairs, this analysis uses a feature extraction algorithm on the PMT waveforms, which reconstructs the constituent PMT pulses caused by individual photo electrons. 
In a second step a Multivariate Analysis is adopted to reduce the background further.

The variables used for background reduction are:

\renewcommand{\labelitemi}{$\bullet$}
\begin{enumerate}
\item The event duration $\Delta{\mathrm{t}}$ defined as the time difference between the last and first pulse registred by a DOM in an event. 
\item The reconstructed speed v from the linefit.
\item The number of clusters ($N_{\mathrm{clusters}}$), which 
is defined by the reconstructed pulses on all DOMs sorted into
groups of pulses which occur close in space and time. Each such group is
called a cluster and the total number of these clusters in an event is
used as a cut variable. 
Bright signal tracks tend to have a higher number of clusters than atmospheric muon background events. 
\item The total number of photoelectrons collected in the whole detector divided by the event duration, $Q_{\mathrm{tot}} / \Delta{\mathrm{t}}$.

\item Median of the distance between clusters along the reconstructed track.

\item The center of gravity (COG) of the event, defined as the average spatial coordinates of all hits.
\suspend{enumerate}

Straight cuts are applied to variables 1-5, chosen to substantially reduce background
while keeping the signal efficiency reasonably high 
(Table \ref{table:cuts_effic_ic59}). The cut on variable 6 removes events that only traverse  
 a corner of the detector.

After applying those cuts, a Multivariate Analysis is performed on the remaining data to define a final selection criterion. 
In addition to variables 3, 4 and 5 this Multivariate Analysis considers the following variables:

\renewcommand{\labelitemi}{$\bullet$}
\resume{enumerate}
\item Mean distance of the hit DOMs to the center of gravity (COG) of the event divided by the event duration.
\item Number of clusters divided by the event duration.
\item Number of simple multiplicity triggers divided by the number of strings with hit DOMs
\end{enumerate}

\subsection{Signal Expectations}
Data are divided into two sets according to the monopole track brightness (i.e. the catalysis cross section). 
The cross section values for which we optimized the analysis and derive flux limits
are  $\sigma_{\mathrm{cat}} = 1.7 \cdot 10^{-22} \mathrm{cm^{2}}$ and 
$\sigma_{\mathrm{cat}} = 1.7 \cdot 10^{-23} \mathrm{cm^{2}}$, which correspond to $\lambda_{\mathrm{cat}}=1\,\mathrm{mm}$ and $\lambda_{\mathrm{cat}}=1\,\mathrm{cm}$ respectively.

\begin{figure}
	\centering
	\includegraphics[width=0.9\columnwidth]{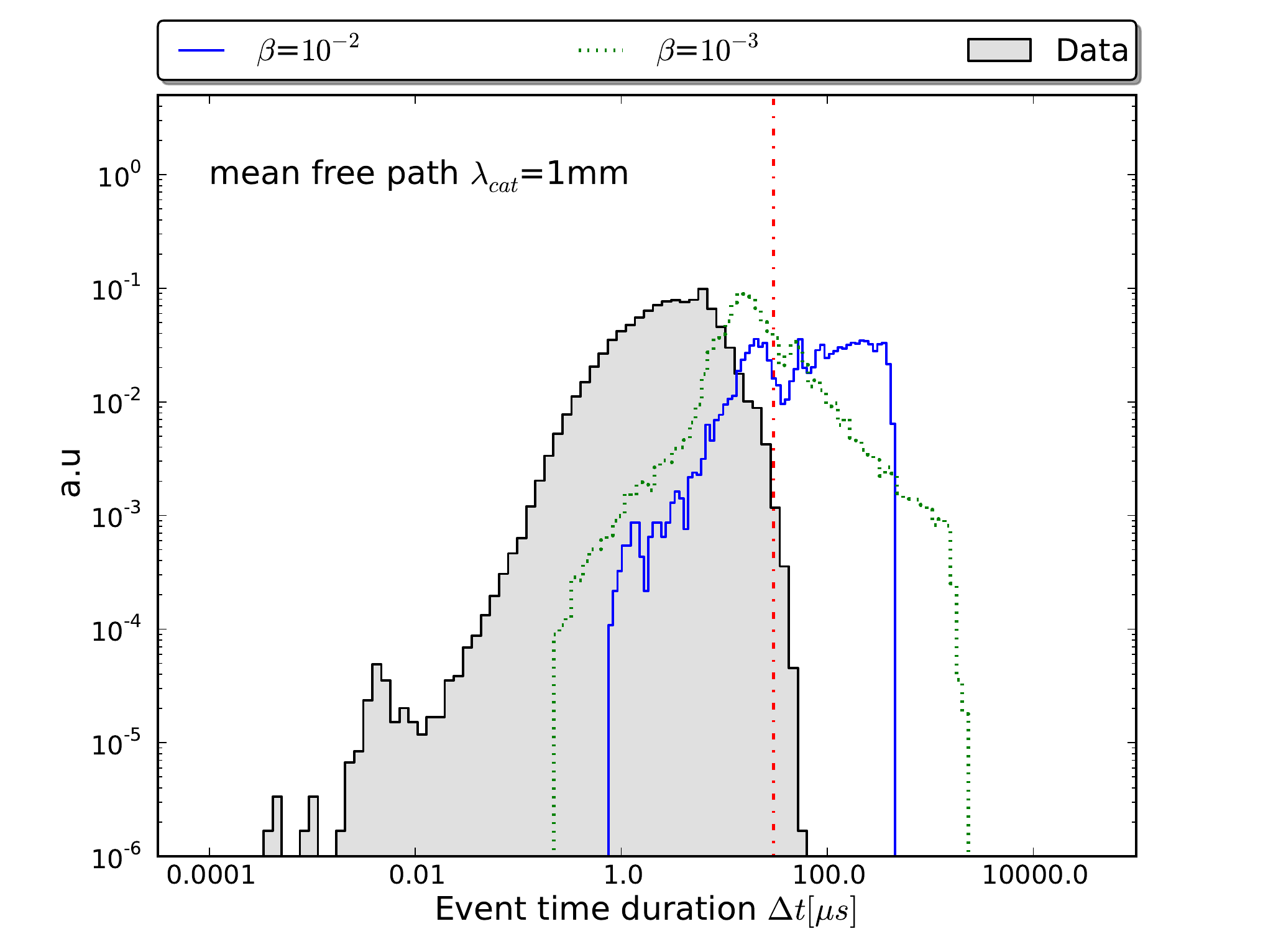}\\\vspace{0.25cm}
	\includegraphics[width=0.9\columnwidth]{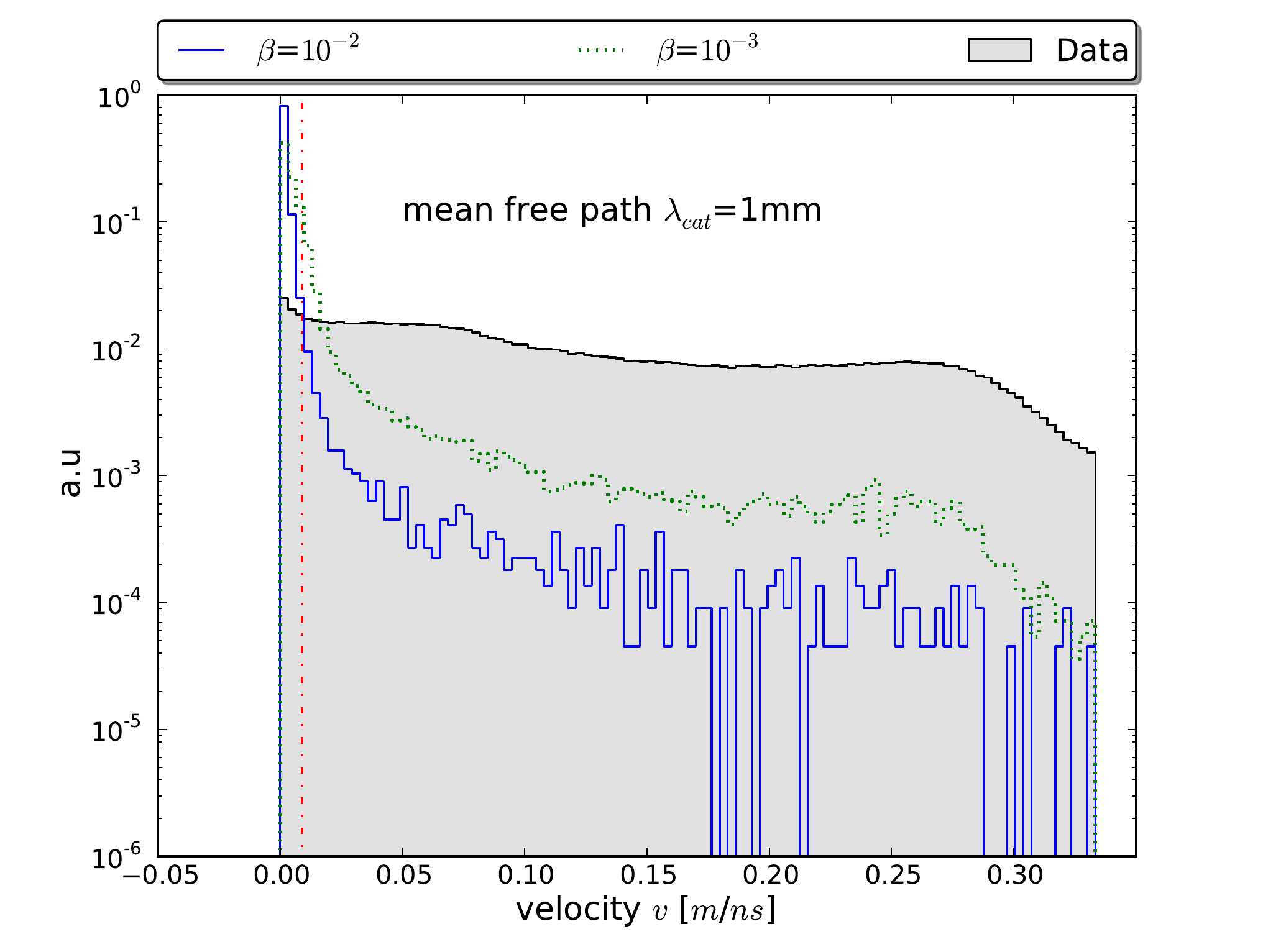}
	\caption{Top: Distribution of the event duration $\Delta{\mathrm{t}}$, for experimental data and simulated bright monopoles with  $\lambda_{\mathrm{cat}}$ = 1\,mm (i.e. $\sigma_{\mathrm{cat}} =1.7\cdot 10^{-22} \mathrm{cm^{2}}$), before applying a cut. The green histograms represents
     monopoles with $\beta =10^{-3}$, the blue histogram with $\beta =10^{-2}$. The gray
     histogram represents the data. The red dashed line marks the value of the chosen cut which is set at $\Delta{\mathrm{t}} > 30\,\mu s$. 
     Bottom: The same for the reconstructed speed v with a cut at v < $9\cdot 10^{-3}$\,m/ns. Histograms are normalized to 1.}
\label{fig:time_durations_velocity_1}
\end{figure}

Figure \ref{fig:time_durations_velocity_1} compares event duration $\Delta{\mathrm{t}}$
and reconstructed speed v of experimental data 
to those of bright monopoles with simulated $\lambda_{\mathrm{cat}}$ = 1\,mm and 
speeds $\beta$ of $10^{-2}$ and $10^{-3}$. The signal efficiencies are 57.0\% and 75.4\% at the filter level for $\beta =10^{-2}$ and $\beta=10^{-3}$, respectively. A cut $\Delta{\mathrm{t}} > 30\,\mu$s reduces the data by a factor $8 \cdot 10^{-4}$ while keeping 43.5\% of the signal for $\beta = 10^{-2}$ and 45.3\% for $\beta = 10^{-3}$ at the filter level. Note that the average duration of triggered events is shorter for slower speeds than for faster, because slower monopole events are more likely to be split into multiple sub-events.

Relativistic single-muon tracks have a reconstructed speed  v around $0.3\,\mathrm{m/ns}$. 
Events having
passed the preceding cut on the  time duration $\Delta{\mathrm{t}}$ are enriched with coincident muons from uncorrelated
air showers, resulting in a lower v (see Fig. \ref{fig:time_durations_velocity_1}, bottom).
For monopoles, the speed v is close to the simulated values. 
Cutting at v < $9\cdot 10^{-3}$\,m/ns (corresponding to $\beta < 3\cdot 10^{-2}$) reduces
the background by another order of magnitude.

Further cuts on variables 3-6 reduce the background by another factor two. 
In total after this first set of cuts the data rate is reduced by a factor $5\cdot 10^{-5}$ 
while the signal efficiencies only drop to 33.6\% and 34.3\% for $\beta=10^{-2}$ and $\beta=10^{-3}$, respectively. 
Data reduction factors, rates and signal efficiencies before and after each applied cut are presented in Table \ref{table:cuts_effic_ic59}.   

\begin{figure}
	\centering
	\includegraphics[width=0.9\columnwidth]{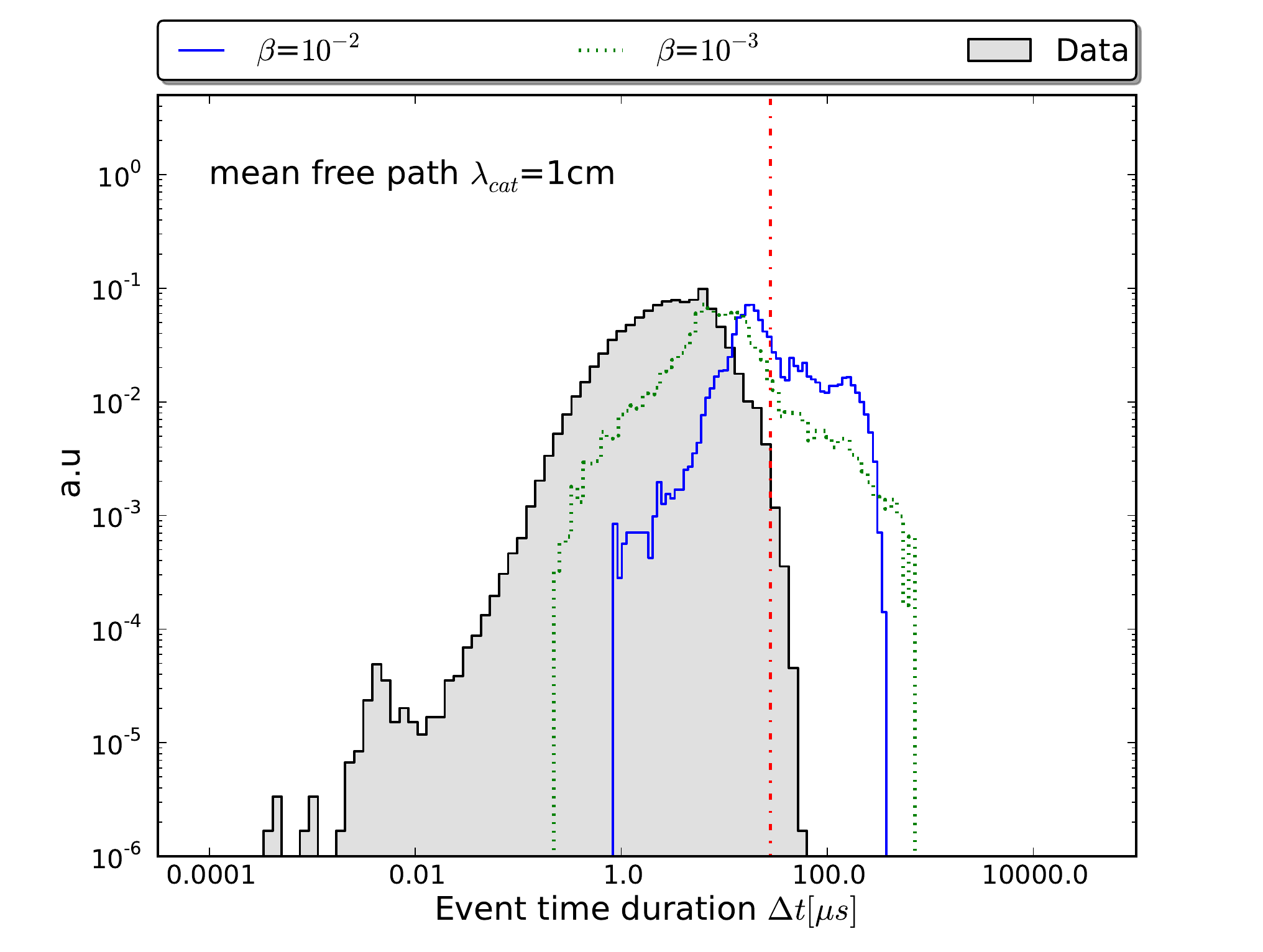}\\\vspace{0.25cm}
	\includegraphics[width=0.9\columnwidth]{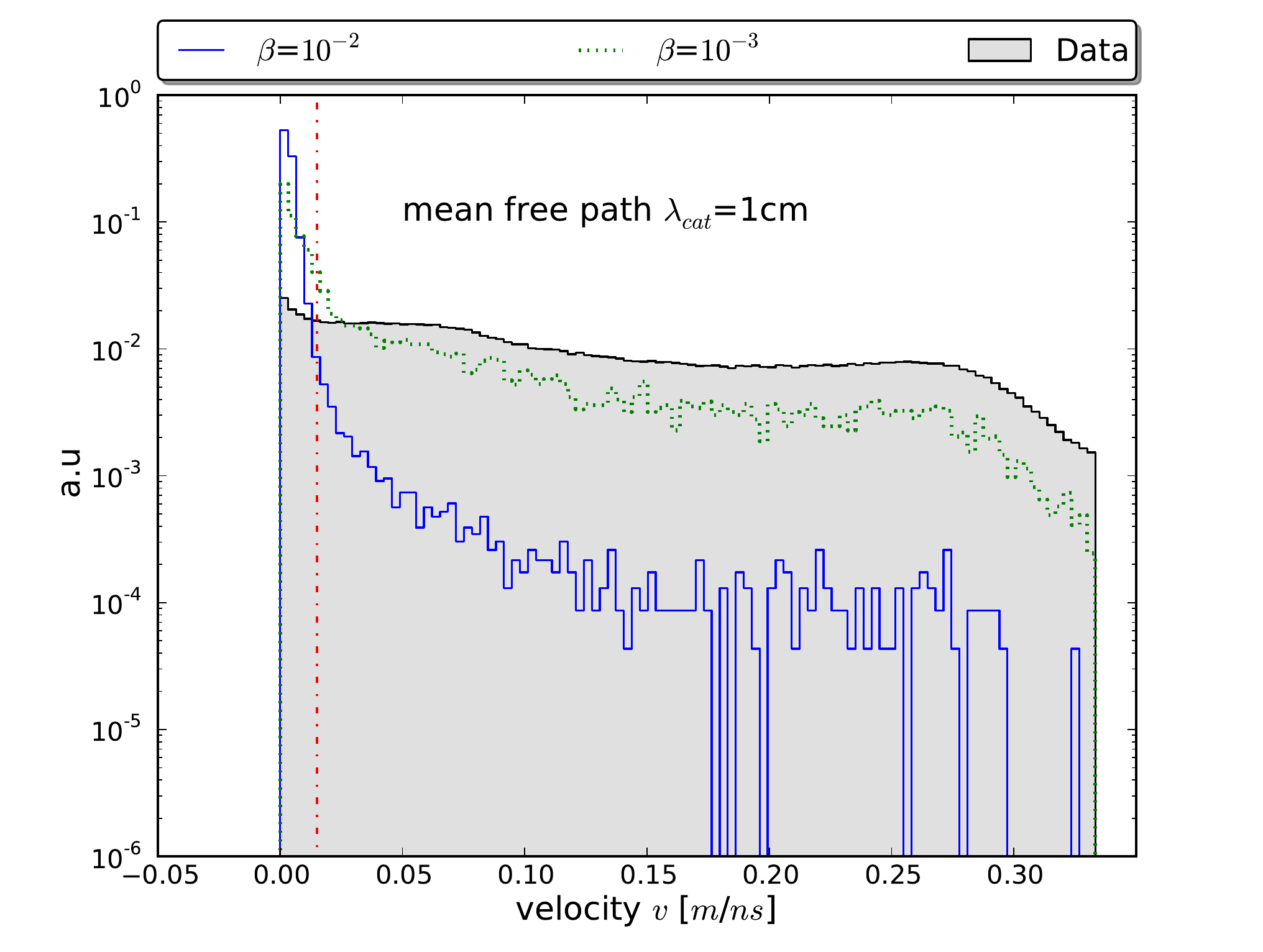}
    \caption{Same as Fig. \ref{fig:time_durations_velocity_1} but with $\lambda_{\mathrm{cat}}=1\,\mathrm{cm}$, i.e. 10 times lower. The red dashed line marks the value of the chosen cut which is set at $\Delta{\mathrm{t}} > 28\,\mu s$ for the event duration and v < $1.5\cdot 10^{-2}$\,m/ns for the speed. Histograms are normalized to 1.}
\label{fig:time_durations_velocity_2}
\end{figure}

For a ten times lower $\sigma_{\mathrm{cat}}$, the monopole tracks are dimmer and the signal efficiency drops dramatically.
Before applying any cut, the efficiencies at the filter level are 41\% and 43\% for $\beta =10^{-2}$ and $\beta =10^{-3}$, respectively.
Figure \ref{fig:time_durations_velocity_2} compares the same variables presented in Fig. \ref{fig:time_durations_velocity_1}. 
Signal and background are much less separated than for $\lambda_{\mathrm{cat}}=1\,\mathrm{mm}$. Moreover,
none of the events with $\beta = 10^{-3}$ has a duration that exceeds 800\,$\mu$s, which is much less than the 3\,ms
necessary to cross the full array; i.e. most events are split into one or more sub-events which have a
shorter event duration in comparison to $\lambda_{\mathrm{cat}}=1\,\mathrm{mm}$. 
The properties of these sub-events are determined by hits from 
muons and from noise falling in the time window of the monopole passage.
Thus the cuts applied on $\Delta{\mathrm{t}}$ and v had to be slightly relaxed 
compared to $\lambda_{\mathrm{cat}}=1\,\mathrm{mm}$: $\Delta{\mathrm{t}}$ > 28 $\mu s$ and v < $1.5\cdot 10^{-2}$\,m/ns.
After excluding events with a reconstructed center of gravity (COG) at outer strings, the data rate is reduced by a factor $3.45\cdot 10^{-4}$. 
Table \ref{table:cuts_effic_ic59} shows the data reduction factors, rates and the final signal efficiencies before and after each applied cut.
The signal efficiencies drop to 13.9\% and 3.1\% for $\beta=10^{-2}$ and $\beta=10^{-3}$ respectively.

\begin{table*}
\centering
\caption{Signal efficiencies, data reduction factors and data rates before and after each cut for both $\sigma_{\mathrm{cat}}$( $\lambda_{\mathrm{cat}}$). For $\sigma_{\mathrm{cat}}=1.7\cdot 10^{-22}\mathrm{cm^{2}}$ the corresponding applied cuts are: "Cut 1 to Cut 6" which are described in subsection 4.2. For $\sigma_{\mathrm{cat}}=1.7\cdot 10^{-23}\mathrm{cm^{2}}$ the applied cuts are: Cut 1, Cut 2 and Cut 6.}
\label{table:cuts_effic_ic59}
\begin{tabular}{l|lllllll}
& & \multicolumn{2}{c}{$\sigma_{\mathrm{cat}}=1.7\cdot 10^{-22}\mathrm{cm^{2}}$, $\lambda_{\mathrm{cat}}=1\,\mathrm{mm}$}\\ 

& Before the cut & Cut 1 & Cut 2 & Cut 3 & Cut 4 & Cut 5 & Cut 6  \\
\hline
$\beta=10^{-2}$   & 57\% & 43.5\% & 42.3\% & 41.9\% & 41.9\% & 41.8\% & 33.6\%       \\
$\beta=10^{-3}$   & 75.4\% & 45.3\% & 41.1\% & 41\%  &41\%    &39.8\%  & 34.3\%      \\
Experiment: reduction factor     &--   & $8 \cdot 10^{-4}$  &  $7.8 \cdot 10^{-5}$ & $6.6 \cdot 10^{-5}$ & $5.7 \cdot 10^{-5}$ & $5.3 \cdot 10^{-5}$ & $4.8 \cdot 10^{-5}$     \\
Experiment: rate[s$^{-1}$]   & 85.5    & $6.8 \cdot 10^{-2}$  &  $6.7 \cdot 10^{-3}$ & $5.6 \cdot 10^{-3}$ & $4.9 \cdot 10^{-3}$ & $4.5 \cdot 10^{-3}$ & $4.1 \cdot 10^{-3}$     \\
\hline
\multicolumn{2}{c}{} \\
& & \multicolumn{2}{c}{$\sigma_{\mathrm{cat}}=1.7\cdot 10^{-23}\mathrm{cm^{2}}$, $\lambda_{\mathrm{cat}}=1\,\mathrm{cm}$} \\

& Before the cut & Cut 1  & Cut 2 & Cut 6 \\
\hline
$\beta=10^{-2}$   & 41\%  & 17.2\% & 17\% & 13.9\%       \\
$\beta=10^{-3}$   & 43\%  &  3.5\% & 3.23\%  & 3.1\%    \\
Experiment: reduction factor & --       & $1.4 \cdot 10^{-3}$& $3.9 \cdot 10^{-4}$ & $3.45 \cdot 10^{-4}$   \\
Experiment: rates[s$^{-1}$]  & 85.5      & $1.2 \cdot 10^{-1}$& $3.3 \cdot 10^{-2}$ & $2.95 \cdot 10^{-2}$   \\
\hline
\end{tabular}
\end{table*}

\subsection{IC-59 Final Cut Optimization}

To optimize the sensitivity for bright monopoles a Multivariate Analysis is used. It classifies each event by a Boosted Decision Trees (BDT) score in the range $[-1,+1]$ \cite{Hocker:TMVA, BDT}.
A BDT score of $-1$ characterizes a background-like event whereas a BDT score of $+1$ characterizes a signal-like event.

For the analysis, a sample of 10\% of all experimental data (burn sample) was divided into two equally-sized sets.
BDTs have been trained on each combination ($\beta$,$\lambda_{\mathrm{cat}}$) of the signal Monte Carlo and on 50\% of the corresponding burn sample, using combinations of the variables described above. The sensitivity was estimated from the other 50\% of the burn sample by fitting an exponential function to the tail of the BDT score distribution.
Over a large range of the BDT scores, the fit describes the data rather well. Still, its extension into the signal region has no 
strict physics justification. 

The final cut on the BDT scores for each combination of ($\beta$,$\lambda_{\mathrm{cat}}$)
is obtained by using the Model Rejection Factor (MRF) method \cite{Hill:MRF}.  
For the chosen high catalysis cross sections
the limits for three ($\beta$, $\lambda_{\mathrm{cat}}$) combinations are significantly better
or comparable to those of the IC-86/DeepCore analysis. The fourth combination ($\beta=10^{-3}$, $\lambda_{\mathrm{cat}}=1$\,cm)
is not competitive because the optimal cut results in 42 expected background events for the full data sample.

\subsection{Results}
\begin{figure}
	\centering
	\includegraphics[width=1.1\columnwidth]{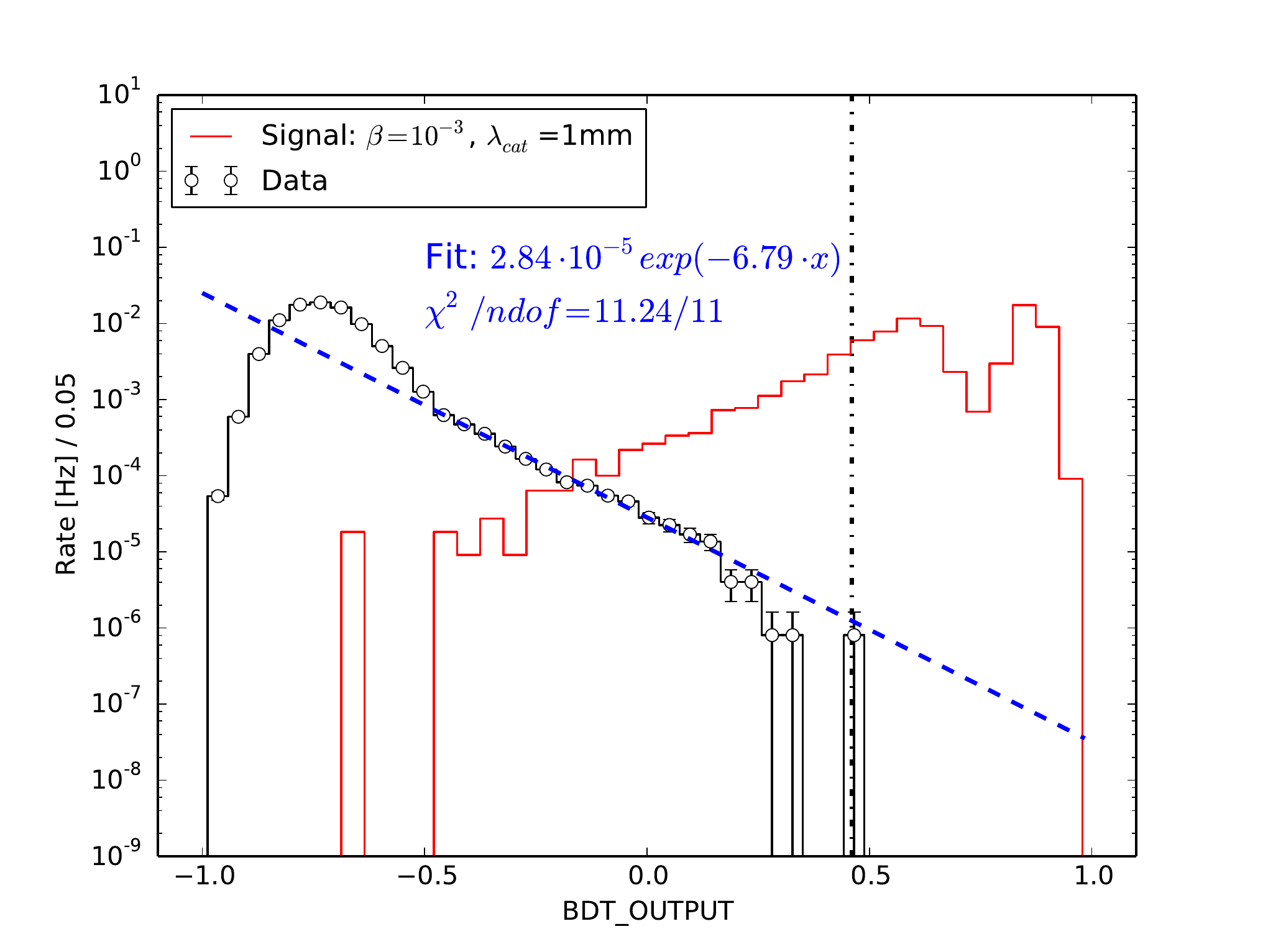}
	\caption{Distribution of the BDT scores, after unblinding, for data  and signal with $\lambda_{\mathrm{cat}}=1$\,mm, and speed $\beta =10^{-3}$. The dot dashed line shows the optimized cut on the BDT score obtained from the Model Rejection Factor method. One event survived the BDT cut and is compatible with background.}
	\label{fig:BDT_limits}
\end{figure}

Figure \ref{fig:BDT_limits} shows the BDT scores for data and signal 
($\beta=10^{-3},\lambda_{\mathrm{cat}}=1$\,mm) for one year data taking (311.25 days live time) 
after the unblinding. The optimized cut on the BDT scores leaves only one event  which merely passes the
cut. No events pass the cuts for $\beta=10^{-2}, \lambda_{\mathrm{cat}}=1$\,mm and 
$\beta=10^{-2}, \lambda_{\mathrm{cat}}=1$\,cm.
The one surviving event was inspected visually. It contains two nearly vertical high-energy muons 
which subsequently cross the whole detector and trigger two neighboring strings. 
It has a time duration $\delta t = 63.6\,\mathrm{\mu s}$ and a reconstructed speed of 
v = $8.5\cdot 10^{-3}$\,m/ns.

The expected number of background events after unblinding is calculated using the exponential fit to the BDT score distributions of the one year experimental data.
The numbers of expected background events and of observed events, as well as the cut values on the BDT score 
for each parameter combination ($\beta$, $\lambda_{\mathrm{cat}}$) are shown in Table \ref{tab:background_events}. 
The higher number of expected background events compared to the actually observed number of events suggests that the exponential 
fit over-estimates the background rather than under-estimates it. 

 \begin{table}
\centering
\begin{tabular}{c|lll}
&\multicolumn{3}{c}{$\sigma_{\mathrm{cat}}=1.7\cdot 10^{-22}\mathrm{cm^{2}}$, $\lambda_{\mathrm{cat}}=1\,\mathrm{mm}$}\\ 
&$BDT_{\mathrm{cut}}$ & $N_{\mathrm{expected}}$ & $N_{\mathrm{observed}}$ \\
\hline
$\beta=10^{-2}$    & 0.46 & $0.6^{+0.2}_{-0.1}$  & 0       \\
$\beta=10^{-3}$    & 0.48 & $4.8^{+0.7}_{-0.6}$  & 1      \\
\hline
\multicolumn{4}{c}{} \\
&\multicolumn{3}{c}{$\sigma_{\mathrm{cat}}=1.7\cdot 10^{-23}\mathrm{cm^{2}}$, $\lambda_{\mathrm{cat}}=1\,\mathrm{cm}$} \\
&$BDT_{\mathrm{cut}}$ & $N_{\mathrm{expected}}$ & $N_{\mathrm{observed}}$ \\
\hline
$\beta=10^{-2}$ & 0.5 & $3.0^{+0.6}_{-0.5}$ & 0       \\
$\beta=10^{-3}$ & not sensitive & not sensitive & not sensitive    \\

\hline
\end{tabular}
\caption{Number of expected and observed events per year for every ($\beta$, $\lambda_{\mathrm{cat}}$) parameter combination. $N_{\mathrm{expected}}$ is derived from the integral of the fitted BDT scores with an exponential. The integral ranges are from the BDT cut value to unity. The errors on the number of expected background event are 1$\sigma$ errors derived from a toy Monte Carlo experiment.}
\label{tab:background_events}
\end{table}   

\section{Systematic Uncertainties\label{sec:systematics}}

The calculation of upper flux limits takes into account 
the statistical and systematic uncertainties in the background and 
signal predictions. Because the number of expected background
events is estimated from experimental data, only the statistical 
uncertainties  of the fit parameters of the background model are relevant.

For signal the imperfect detector description is taken into account. For example in case of the IC-86/DeepCore search, the random noise leads to an
increase of $n\mathrm{-triplet}$ for signal events.
Furthermore the optical light detection efficiency is important.
This efficiency takes into account the cumulative effect of the light yield of nucleon decays, where a single electromagnetic cascade is simulated instead of several daughter particles, 
the light propagation through the ice and its detection by the DOMs.
These effects result in an uncertainty of the detection efficiency for magnetic monopoles which is used to derive the upper limits.

The impact of these uncertainties on the flux limits is estimated by simulating monopoles with simulation parameters changed within their estimated uncertainties.
The uncertainties of the superimposed background noise, the light yield of nucleon decays and the light propagation through ice are estimated
by their differences in the detection efficiencies of signal simulations taking into account different approaches (Sec. \ref{sec:simulation_monopole}).
For the IC-86/DeepCore analysis the superimposed noise can be described by random and correlated noise hits from experimental data or
noise simulated as a Poisson process and atmospheric muons simulated using the software package CORSIKA.
Since for the IC-59 analysis no unbiased experimental data exists the background noise can be simulated by a noise generator that also takes into account correlated noise hits.
For reasons of simplification the proton decay is simulated as a single electromagnetic cascade with an isotropic direction which is valid as long as the mean free path is much smaller than the IceCube spacing.
Due to kinematics in the proton decay (Eq. \ref{eqn:promising_decay_channel}) two back-to-back electromagnetic cascades with an isotropic direction have to be simulated.
The uncertainties due to this simplification are estimated by the differences between both approaches.
For the light propagation through ice the two ice models described in \cite{IceCube:IceModel_AHA} and \cite{IceCube:SPICEMie} are used.
The uncertainty of the optical efficiency of DOMs can be estimated as $\pm 10\%$.
Signal simulations based on optical DOM efficiencies varied by $\pm 10\%$ are compared with simulations based on the default settings.
The differences in the detection efficiencies are used as an estimate for the uncertainty.

We quantify each systematic effect $i$  by the  ratio of the resulting 
detection efficiency $\epsilon_{\mathrm{i}}$ relative to the detection efficiency with baseline assumptions $\epsilon_{0}$:

\begin{equation}
R_{\mathrm{i}}\left(\beta,\sigma_{\mathrm{cat}}\right) = \frac{\epsilon_{\mathrm{i}}\left(\beta,\sigma_{\mathrm{cat}}\right)}{\epsilon_{0}\left(\beta,\sigma_{\mathrm{cat}}\right)} ~.
\end{equation}
The resulting changes are displayed in Table \ref{tab:sys_eff_area}.

\begin{table*}
  \centering
  \caption{The impact of different systematic uncertainties on the detection efficiencies of magnetic monopoles depending on the mean free path $\lambda_{\mathrm{cat}}$ and the monopole speed $\beta$. The first column shows the impact of different assumptions for the superimposed background noise. Also the uncertainties of the simplified nucleon decay simulation (second column), the optical DOM efficiency (third column) and the optical ice properties (fourth column) are shown.}
\label{tab:sys_eff_area}
\begin{tabular}{cr|ll|ll|ll|ll}
     & & \multicolumn{2}{c|}{Noise simulation}&  \multicolumn{2}{c|}{Nucleon decay simulation} & \multicolumn{2}{c|}{Optical DOM efficiency} & \multicolumn{2}{c}{Optical ice properties} \\
     & $\lambda_{\mathrm{cat}}\;[\mathrm{m}]$ & \multicolumn{1}{|c}{$\beta=10^{-2}$} & \multicolumn{1}{c|}{$\beta=10^{-3}$} & \multicolumn{1}{|c}{$\beta=10^{-2}$} & \multicolumn{1}{c|}{$\beta=10^{-3}$} & \multicolumn{1}{|c}{$\beta=10^{-2}$} & \multicolumn{1}{c|}{$\beta=10^{-3}$} & \multicolumn{1}{|c}{$\beta=10^{-2}$} & \multicolumn{1}{c}{$\beta=10^{-3}$} \\
     \noalign{\smallskip}\hline\noalign{\smallskip}
     \multirow{7}{*}{IC-86} & 3.0 & +29\%$/$-2\% & +23\% & +36\% & -16\% & +36\%$/$-23\% & +46\%$/$-29\% & +20\%$/$-11\% & +72\% \\
     & 1.0 & +52\% & +6\%$/$-11\% & +27\% & +1\%$/$-11\% & +14\%$/$-7\% & +16\%$/$-20\% & -17\% & +12\% \\
     & 0.3 & +31\% & +5\%$/$-8\% & +15\% & +1\%$/$-8\% & $\pm$10\% & $\pm$11\% & -16\% & +8\%$/$-2\% \\
     & 0.1 & +19\%$/$-1\% & +3\%$/$-4\% & $\pm$7\% & +1\%$/$-6\% & +5\%$/$-11\% & +8\%$/$-6\% & -15\% & +1\%$/$-6\% \\
     & 0.03 & +17\%$/$-2\% & +1\%$/$-5\% & +9\%$/$-4\% & +1\%$/$-4\% & +10\%$/$-5\% & +6\%$/$-3\% & -9\% & -7\% \\
     & 0.01 & +15\%$/$-4\% & -4\% & +11\%$/$-2\% & $\pm$2\% & +12\%$/$-1\% & +5\%$/$-0.3\% & $\pm$5\% & -10\% \\
     & 0.001 & +15\%$/$-4\% & -4\% & +11\%$/$-2\% & $\pm$2\% & +12\%$/$-1\% & +5\%$/$-0.3\% & $\pm$ 5\% & -10\% \\
   \noalign{\smallskip}\hline\noalign{\smallskip}
    \multirow{2}{*}{IC-59} & 0.01 & +2\% & -- & -5\% & -- & +9\%$/$-6\% & -- & +7\% & -- \\
     & 0.001 & $-3\%$ & $+5\%$ & -3\% & -1\% & -5\% & +9\%$/$-2\% & +2\% & +4\% \\
   \noalign{\smallskip}\hline 
  \end{tabular}
\end{table*}

Note, that these calculations  are limited by computing resources and correspondingly by the statistics of simulated events.
The resulting statistical uncertainties of effective areas 
are typically a few percent as shown in Table \ref{tab:stat_eff_area}
and included in the total error.

\begin{table}
  \centering
  \caption{The statistical uncertainties of the calculated effective areas
for different  mean free path $\lambda_{\mathrm{cat}}$ and  speed $\beta$ for  the IC-86 and IC-59 analyses.}
  \label{tab:stat_eff_area}
  	 \begin{tabular}{rr|ll}
    & &\multicolumn{2}{|c}{Statistical uncertainties}  \\
    &$\lambda_{\mathrm{cat}} [m]$ & $\beta = 10^{-2}$ & $\beta = 10^{-3}$  \\
    \noalign{\smallskip}\hline\noalign{\smallskip}
     \multirow{7}{*}{IC-86} & 3.0 & $9\%$ & $10\%$ \\
    & 1.0 & 8\% & 5\% \\
    & 0.3 & 8\% & 2\% \\
    & 0.1 & 8\% & 2\% \\
    & 0.03 & 9\% & 2\% \\
    & 0.01 & 10\% & 2\% \\
    & 0.001 & 12\% & 2\% \\
    \noalign{\smallskip}\hline
    \multirow{2}{*}{IC-59} & 0.01 & 0.2\% & -- \\
    & 0.001 & 0.1\% & 0.1\% \\
     \noalign{\smallskip}\hline 
  \end{tabular}
\end{table}

For the calculation of the final flux limits we perform high statistics computer experiments. In each we randomize 
the effect of each systematics  effect $R_{\mathrm{i}}$
according to its specific uncertainty. 
For each parameter combination $\beta$ and $\sigma_{\mathrm{cat}}$ 
this results in the effective probability density distribution for the relative change
of the detection efficiency $R$ taking into account all uncertainties.

\begin{figure}
    \centering
    \includegraphics[width=0.9\columnwidth]{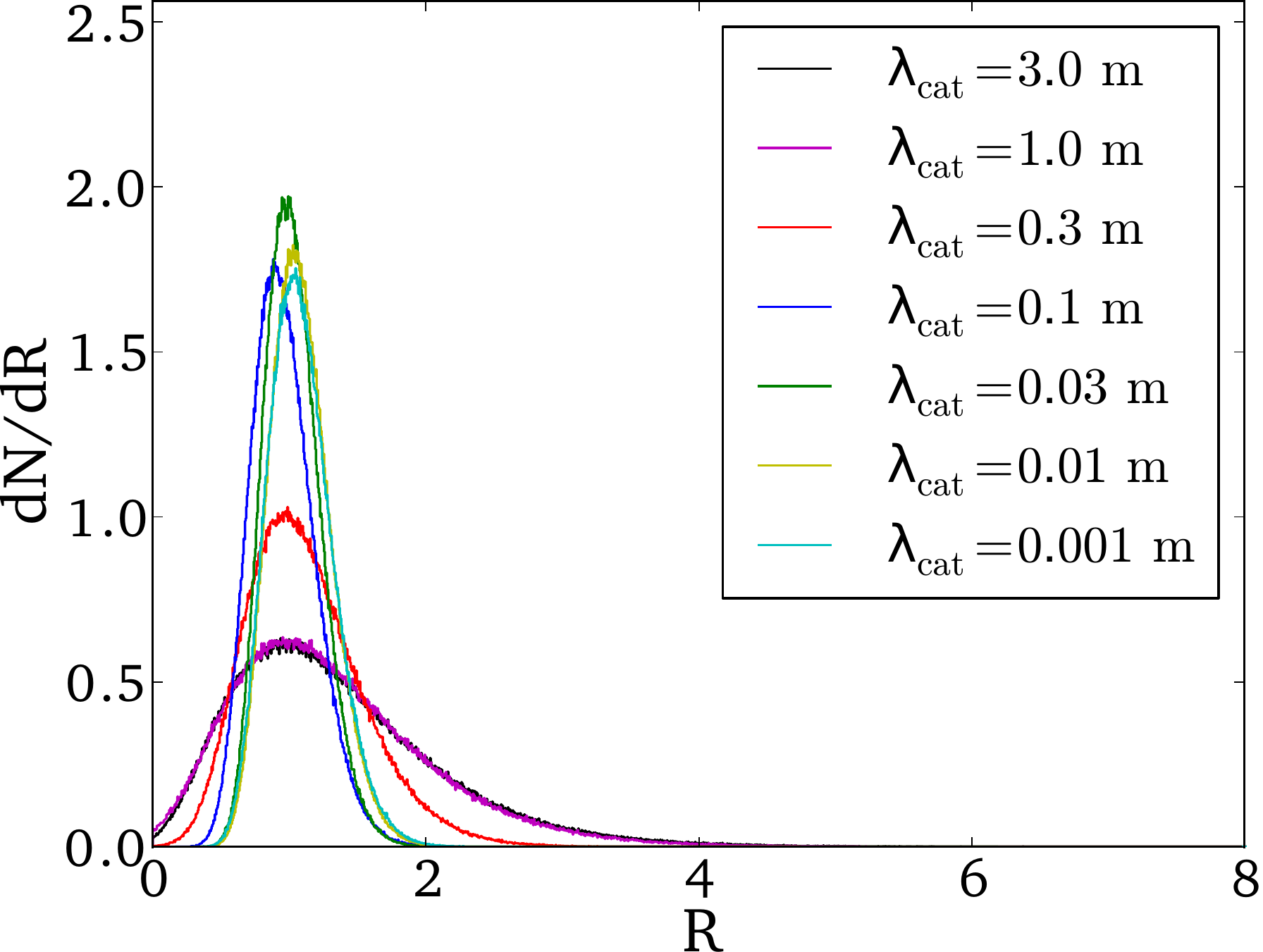}
    \caption{Probability density distribution for systematic signal uncertainties for $\beta=10^{-2}$ monopoles for the IC-86/DeepCore analysis.}\label{fig:signal_systematics}
\end{figure}

An example is shown in Fig. \ref{fig:signal_systematics}. Multiplying these distributions by the detection efficiency
with baseline assumptions $\epsilon_0$ one gets the probability density distributions for the detection efficiencies.

\section{Flux Limits}\label{sec:flux_limits}

The flux limits on non-relativistic magnetic monopoles are calculated
assuming an isotropic flux and the proton decay channel $p\to e^+ \pi^0 $ (Eq. \ref{eqn:promising_decay_channel}) with the catalysis cross section $\sigma_{\mathrm{cat}}$ which depends on the speed $\beta$ (Eq. \ref{eqn:sigma_cat_theo}).
Using the quantity $\hat{l}$ (Eq. \ref{eqn:monopole_photon_yield}) the flux limits can also be expressed without assuming a specific decay channel.

The flux limits are calculated based on a generalization of the approach by Rolke et al. \cite{RolkeLimits},
which takes into account the uncertainties of the signal detection efficiency and the expected number of background events.
Therefore, a three-dimensional likelihood fit is performed with the following parameters:
expected number of signal events $\mu$, signal detection efficiency $\epsilon$ and expected number of background events $n_{\mathrm{b}}$.
The likelihood function is defined by

\begin{equation}
    L(\mu,\epsilon,n_{\mathrm{b}} \mid n_{\mathrm{obs}}) = \frac{\lambda^{n_{\mathrm{obs}}} \cdot \mathrm{e}^{-\lambda}}{n_{\mathrm{obs}}!} \cdot f_{\mathrm{s}}(\epsilon) \cdot f_{\mathrm{b}}(n_{\mathrm{b}}),
\end{equation}
where the number of observed events $n_{\mathrm{obs}}$ follows a poisson distribution with the expectation value $\lambda = \epsilon \mu + n_{\mathrm{b}}$.
The functions $f_{\mathrm{s}}(\epsilon)$ and $f_{\mathrm{b}}(n_{\mathrm{b}})$ represent the probability density distributions
of the signal detection efficiency and the expected number of background events (Figs. \ref{fig:pdf_background} and \ref{fig:signal_systematics}).

The flux limits for each monopole speed $\beta$ and catalysis cross section $\sigma_{\mathrm{cat}}$ are calculated by

\begin{equation}
\Phi_{90}\left(\beta,\sigma_{\mathrm{cat}}\right) = \frac{\mu_{90}}{A_{\mathrm{gen}} \cdot t \cdot \Omega},
\end{equation}
where $\mu_{90}$ is the upper limit at the 90\% confidence level for $n_{\mathrm{obs}}=1$ event. The upper limit is obtained from the profile likelihood function defined in \cite{RolkeLimits}.
In order to exclude unphysical values of $\mu_{90}$ the expected number of signal events $\mu$ is constrained to be greater than or equal to zero. As a consequence of this method the upper limits at the 90\% confidence level have a slight over-coverage of about 5\%. The other parameters are the size of the signal generation disc $A_{\mathrm{gen}}$, the solid angle $\Omega = 4 \pi \mathrm{sr}$ and the livetime $t$.

\begin{figure}
	\centering
	\includegraphics[width=1.0\columnwidth]{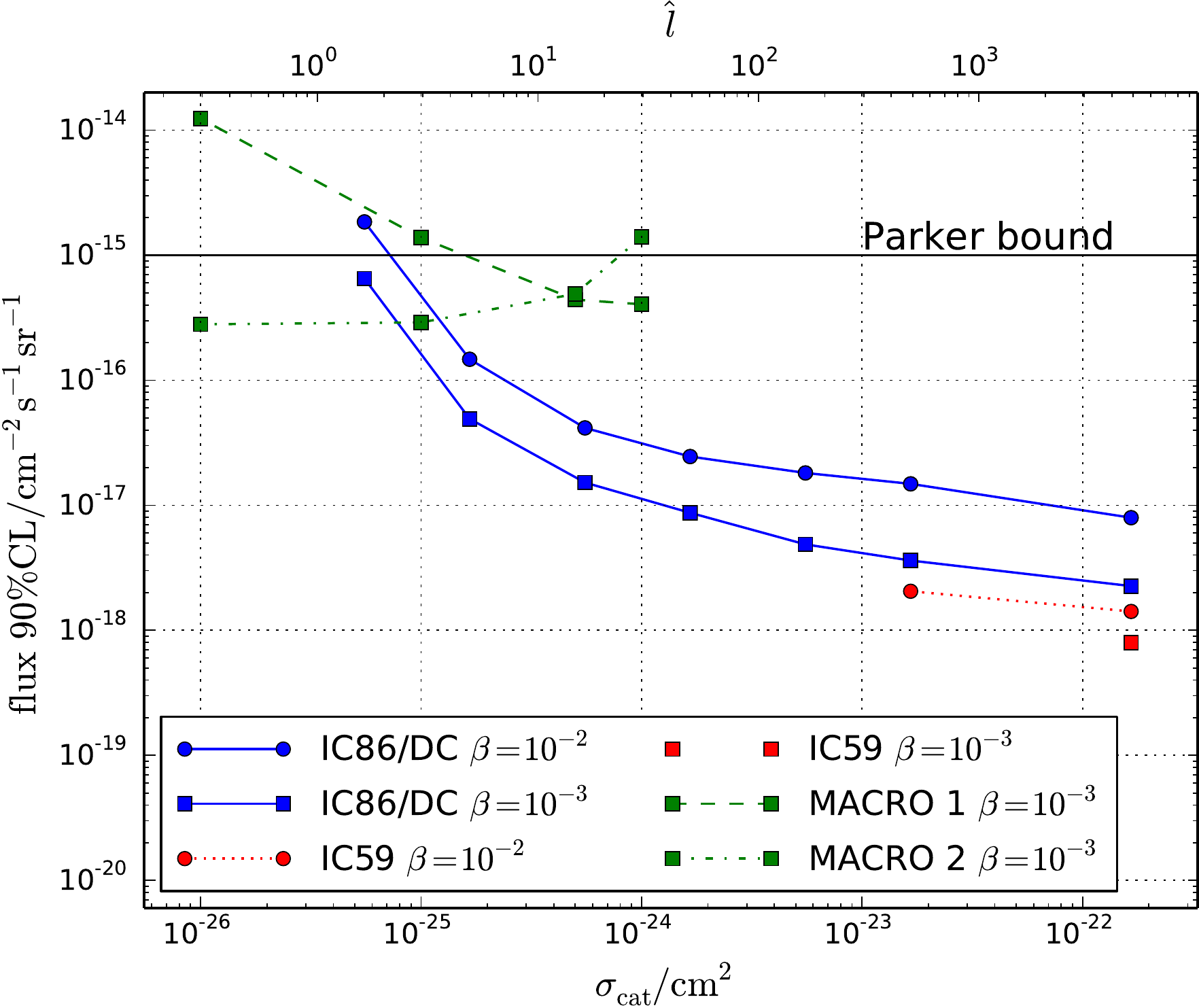}
	\caption{Upper limits on the flux of non-relativistic magnetic monopoles depending on the speed $\beta$ and catalysis cross section $\sigma_{\mathrm{cat}}$ of the IC-59 analysis and IC-86/DeepCore analysis. The dashed lines are limits published by the MACRO experiment \cite{MACRO:SlowMonopoles}. Here, MACRO 1 is an analysis developed for monopoles catalyzing the proton decay. MACRO 2 is the standard-MACRO-analysis, which is sensitive to monopoles ionizing the surrounding matter. Additionally, the IceCube limits are shown as a function of $\hat{l}$ which is proportional to the averaged Cherenkov photon yield per nuceleon decay (not valid for MACRO limits).}
	\label{fig:limits}
\end{figure}

Figure \ref{fig:limits} shows the resulting direct detection limits on the flux of non-relativistic magnetic monopoles in comparison to the current best experimental flux limits by the MACRO experiment \cite{FinalMACRO,MACRO:SlowMonopoles}.

Above  $\sigma_{\mathrm{cat}} = 10^{-25} \, \mathrm{cm}^2$ 
corresponding to $\lambda_{\mathrm{cat}} < 3\,\mathrm{m}$ the previous
flux limits are improved by  more than one order of magnitude. 
Moreover, for such large cross sections 
the monopole flux can be constrained up to a level 
$$\Phi_{90} \le 10^{-18} \mathrm{cm^{-2}s^{-1}sr^{-1}},$$
which is three orders of magnitude below the Parker bound.

Assuming monopoles are the dominant part of Dark Matter, i.e. the relic mass density of monopoles is similar to the Dark Matter mass density,
our most stringent flux limits constrain the monopole mass to be at least of the order of the Planck mass $m_{\mathrm{pl}}=1.22 \cdot 10^{19}\,\mathrm{GeV}$ \cite{EarlyUniverse:Kolb}.
This implies that monopoles with masses significantly smaller than the Planck mass do not contribute dominantly to the Dark Matter mass density.

Indirect searches for monopole induced proton decays set very strong bounds on monopoles with non-relativistic speeds,
e.g. the limits from Super-Kamiokande \cite{indirectSKlimits} assuming gravitational trapping of monopoles in the sun.
Also a variety of bounds based on observations of neutron stars, white dwarfs and gas giants have been obtained.
These bounds range from $\sim 10^{-18}-10^{-29}\,\mathrm{cm}^{-2}\,\mathrm{s}^{-1}\,\mathrm{sr}^{-1}$ 
and depend on the catalysis cross sections as well as on details of the assumed astrophysical scenarios \cite{Harvey:Bounds_neutronstars,Freese:Bounds_whitedwarfs,Arafune:Bounds_jovianplanets}.
Although the direct IceCube searches are not as stringent as indirect searches the former are not affected by astrophysical uncertainties.
Thus the direct IceCube limits can be considered as a robust upper bound on the monopole flux, if the Rubakov-Callan effect is realized in nature.

\section{Summary and Outlook}

Data taken from May 2011 until May 2012 with a dedicated slow-particle trigger and for the brightest monopoles data taken from May 2009 until May 2010 with
standard-IceCube-triggers were analysed. The analysis, which is based on data of the slow-particle trigger, was developed by using simulated monopole events and experimental data to estimate background properties.
For this first analysis of such a signal in IceCube a robust approach based on a single final selection criterion and the comparison between the number of expected background events and observed experimental events is chosen.
Using experimental data, the number of expected background events can be estimated to $n_{\mathrm{b}} = 3.2^{+1.8}_{-1.1}$. 

The IC-59 analysis based on standard-IceCube-triggers is sensitive only for bright monopoles with $\sigma_{\mathrm{cat}} > 1.7 \cdot 10^{-23}\,\mathrm{cm^2}$.
The analysis used Boosted Decision Trees (BDT) to discriminate between monopole signal and background. The expected number of background events is derived from a fit of the BDT scores tails with an exponential function for each ($\beta$,$\lambda_{\mathrm{cat}}$).
The number of observed events after unblinding is $1$ for an expected background of $4.8_{-0.6}^{+0.7}$. This event contains multiple coincident muons, which renders it compatible with a background event. 
The obtained flux limits for $\beta=10^{-2}$ and $\lambda_{\mathrm{cat}} =0.01$\,m, 
$0.001$\,m from the IC-59 analysis are better than the ones from the IC-86/DeepCore analysis because of the bigger effective area. 
For $\beta=10^{-3}$ the limits are comparable since the standard IceCube triggers are less sensitive to the monopole signal in comparison to the 
dedicated slow-particle trigger.

In both analyses no monopole signal has been observed. Thus, the limits on the flux of non-relativistic magnetic monopoles -- catalysing the proton decay -- are improved by about more than one order of magnitude in comparison to MACRO \cite{MACRO:SlowMonopoles} for most of the investigated parameter space and reach down to about three orders of magnitude below the Parker limit.

Since May 2012 the dedicated slow-particle trigger has been updated to the full IceCube detector.
From this upgrade, we expect an improvement in sensitivity by roughly an order of magnitude \cite{IceCube:ICRC2013_slowMPs}.
This gain is supplemented by  improvements of the data selection
which have been developed after completion of this analysis.
Examples are the implementation of a Kalman-filter based  HLC hit selection, 
which  improves the angular and speed 
reconstruction, and the implementation of an event selection based on a Boosted Decision Tree \cite{Zierke:Masterarbeit}.

\begin{acknowledgements}
    We acknowledge the support from the following agencies:
    U.S. National Science Foundation-Office of Polar Programs,
    U.S. National Science Foundation-Physics Division,
    University of Wisconsin Alumni Research Foundation,
    the Grid Laboratory Of Wisconsin (GLOW) grid infrastructure at the University of Wisconsin - Madison, the Open Science Grid (OSG) grid infrastructure;
    U.S. Department of Energy, and National Energy Research Scientific Computing Center,
    the Louisiana Optical Network Initiative (LONI) grid computing resources;
    Natural Sciences and Engineering Research Council of Canada,
    WestGrid and Compute/Calcul Canada;
    Swedish Research Council,
    Swedish Polar Research Secretariat,
    Swedish National Infrastructure for Computing (SNIC),
    and Knut and Alice Wallenberg Foundation, Sweden;
    German Ministry for Education and Research (BMBF),
    Deutsche Forschungsgemeinschaft (DFG),
    L.B. was funded by the DFG Sonderforschungsbereich 676,
    Helmholtz Alliance for Astroparticle Physics (HAP),
    Research Department of Plasmas with Complex Interactions (Bochum), Germany;
    Fund for Scientific Research (FNRS-FWO),
    FWO Odysseus programme,
    Flanders Institute to encourage scientific and technological research in industry (IWT),
    Belgian Federal Science Policy Office (Belspo);
    University of Oxford, United Kingdom;
    Marsden Fund, New Zealand;
    Australian Research Council;
    Japan Society for Promotion of Science (JSPS);
    the Swiss National Science Foundation (SNSF), Switzerland;
    National Research Foundation of Korea (NRF);
    Danish National Research Foundation, Denmark (DNRF)
\end{acknowledgements}

\bibliographystyle{h-physrev}

\begin{thebibliography}{10}

\bibitem{Dirac:1931}
P.~A. Dirac,
\newblock Proc. Roy. Soc. Lond. {\bf A133}, 60 (1931).

\bibitem{Georgi:1974FirstGUT}
H.~Georgi and S.~Glashow,
\newblock Phys. Rev. Lett. {\bf 32}, 438 (1974).

\bibitem{tHooft:1974MonopoleSolution}
G.~'t~Hooft,
\newblock Nucl. Phys. {\bf B79}, 276 (1974).

\bibitem{Polyakov:1974MonopoleSolution}
A.~M. Polyakov,
\newblock JETP Lett. {\bf 20}, 194 (1974).

\bibitem{Georgi:1974_Interaction_GUT}
H.~Georgi, H.~R. Quinn, and S.~Weinberg,
\newblock Phys. Rev. Lett. {\bf 33}, 451 (1974).

\bibitem{Daniel:1979_SU5Monopoles}
M.~Daniel, G.~Lazarides, and Q.~Shafi,
\newblock Nucl. Phys. B {\bf 170}, 156 (1980).

\bibitem{Lazarides}
G.~Lazarides, C.~Panagiotakopoulos, and Q.~Shafi,
\newblock Phys. Rev. Lett. {\bf 58}, 1707 (1987).

\bibitem{Kephart}
T.~W. Kephart and Q.~Shafi,
\newblock Phys. Lett. {\bf B520}, 313 (2001), hep-ph/0105237.

\bibitem{Wick:CosmicFluxMM}
S.~D. Wick, T.~W. Kephart, T.~J. Weiler, and P.~L. Biermann,
\newblock Astropart. Phys. {\bf 18}, 663 (2003), astro-ph/0001233.

\bibitem{Kibble}
T.~Kibble,
\newblock J. Phys. {\bf A9}, 1387 (1976).

\bibitem{Preskill:MM}
J.~Preskill,
\newblock Ann. Rev. Nucl. Part. Sci. {\bf 34}, 461 (1984).

\bibitem{Turner:ParkerBound}
M.~S. Turner, E.~N. Parker, and T.~Bogdan,
\newblock Phys. Rev. {\bf D26}, 1296 (1982).

\bibitem{Groom:SearchSupermassiveMM}
D.~E. Groom,
\newblock Phys. Rept. {\bf 140}, 323 (1986).

\bibitem{ExtensionParkerBound}
F.~C. Adams {\em et~al.},
\newblock Phys. Rev. Lett. {\bf 70}, 2511 (1993).

\bibitem{FinalMACRO}
MACRO Collaboration, M.~Ambrosio {\em et~al.},
\newblock Eur. Phys. J. {\bf C25}, 511 (2002), hep-ex/0207020.

\bibitem{MACRO:SlowMonopoles}
MACRO Collaboration, M.~Ambrosio {\em et~al.},
\newblock Eur. Phys. J. {\bf C26}, 163 (2002), hep-ex/0207024.

\bibitem{icecube:relMPs}
IceCube Collaboration, R.~Abbasi {\em et~al.},
\newblock Phys. Rev. {\bf D87}, 022001 (2013), arXiv/1208.4861.

\bibitem{IceCube:ICRC2013_relMPs}
IceCube Collaboration, M.~Aartsen {\em et~al.},
\newblock p.~9 (2013), arXiv/1309.7007.

\bibitem{IceCubeSensitivity}
IceCube Collaboration, J.~Ahrens {\em et~al.},
\newblock Astropart. Phys. {\bf 20}, 507 (2004), astro-ph/0305196.

\bibitem{Tompkins:CherenkovEmission}
D.~R. Tompkins,
\newblock Phys. Rev. {\bf 138}, B248 (1965).

\bibitem{IceCube:PMT}
IceCube Collaboration, R.~Abbasi {\em et~al.},
\newblock Nucl. Instrum. Meth. {\bf A618}, 139 (2010), arXiv/1002.2442.

\bibitem{icecube:daq}
IceCube Collaboration, R.~Abbasi {\em et~al.},
\newblock Nucl. Instrum. Meth. {\bf A601}, 294 (2009), arXiv/0810.4930.

\bibitem{IceCubePerform}
IceCube Collaboration, A.~Achterberg {\em et~al.},
\newblock Astropart. Phys. {\bf 26}, 155 (2006), astro-ph/0604450.

\bibitem{IceCubeDeepCore}
IceCube Collaboration, R.~Abbasi {\em et~al.},
\newblock Astropart. Phys. {\bf 35}, 615 (2012), arXiv/1109.6096.

\bibitem{Ahlen:modified_BetheBloch}
S.~Ahlen,
\newblock Rev. Mod. Phys. {\bf 52}, 121 (1980).

\bibitem{Kazama:BetheBloch_Correction}
Y.~Kazama, C.~N. Yang, and A.~S. Goldhaber,
\newblock Phys. Rev. {\bf D15}, 2287 (1977).

\bibitem{Bloch:Correction}
F.~Bloch,
\newblock Zeitschrift f\"ur Physik A Hadrons and Nuclei {\bf 81}, 363 (1933).

\bibitem{Ahlen:IonizationSlowMP}
S.~Ahlen and K.~Kinoshita,
\newblock Phys. Rev. {\bf D26}, 2347 (1982).

\bibitem{Ritson:IonizationSlowMP}
D.~Ritson,
\newblock SLAC-PUB-2950  (1982).

\bibitem{Rubakov:protondecay}
V.~Rubakov,
\newblock Nucl. Phys. {\bf B203}, 311 (1982).

\bibitem{CallanJr:protondecay}
J.~Callan, Curtis~G.,
\newblock Nucl. Phys. {\bf B212}, 391 (1983).

\bibitem{DawsonPRD27:GUTdependence}
S.~Dawson and A.~N. Schellekens,
\newblock Phys. Rev. D {\bf 27}, 2119 (1983).

\bibitem{Walsh:GUTdependence}
T.~Walsh, P.~Weisz, and T.~T. Wu,
\newblock Nucl. Phys. B {\bf 232}, 349 (1984).

\bibitem{DawsonPRD28:GUTdependence}
S.~Dawson and A.~N. Schellekens,
\newblock Phys. Rev. D {\bf 28}, 3125 (1983).

\bibitem{Rubakov:GUTdependence}
V.~Rubakov and M.~Serebryakov,
\newblock Nucl. Phys. B {\bf 237}, 329 (1984).

\bibitem{Rubakov:ValuesSigma0}
V.~Rubakov,
\newblock Rept. Prog. Phys. {\bf 51}, 189 (1988).

\bibitem{Arafune:RubakovVelocityDependent}
J.~Arafune and M.~Fukugita,
\newblock Phys. Rev. Lett. {\bf 50}, 1901 (1983).

\bibitem{Nath:sigma_cat}
P.~Nath and P.~Fileviez~Perez,
\newblock Phys. Rept. {\bf 441}, 191 (2007), hep-ph/0601023.

\bibitem{KowalskiCherenkovYield}
{M. Kowalski},
\newblock {\em {Search for Neutrino-Induced Cascades with the AMANDA-II
  Detector}},
\newblock PhD thesis, {HU-Berlin}, {2004},
  {http://edoc.hu-berlin.de/dissertationen/kowalski-marek-paul-2004-01-13/PDF/%
Kowalski.pdf}.

\bibitem{Radel:muonGEANT}
L.~Radel and C.~Wiebusch,
\newblock Astropart. Phys. {\bf 38}, 53 (2012), arXiv/1206.5530.

\bibitem{BeckerSzendyIMB}
R.~Becker-Szendy {\em et~al.},
\newblock Phys. Rev. {\bf D49}, 2169 (1994).

\bibitem{FukugitaKamiokandeII}
M.~Fukugita and A.~Suzuki,
\newblock {\em {Physics and astrophysics of neutrinos}} (Springer, 1994).

\bibitem{DomogatskyGyrlanda-86}
G.~Domogatsky {\em et~al.},
\newblock Present status of baikal deep underwater experiment,
\newblock pp. 737--745, 1986.

\bibitem{BezrukovNT36}
BAIKAL Collaboration, L.~B. Bezrukov {\em et~al.},
\newblock (1995), astro-ph/9601160.

\bibitem{BelolaptikovNT36}
BAIKAL Collaboration, I.~Belolaptikov {\em et~al.},
\newblock Astropart. Phys. {\bf 7}, 263 (1997).

\bibitem{PohlAMANDA}
A.~Pohl,
\newblock {\em "Search for subrelativistic particles with the AMANDA neutrino
  telescope"},
\newblock PhD thesis, 2009, http://wwwiexp.desy.de/groups/astroparticle/pubs/
  Thesis.Arvid.090210.pdf.

\bibitem{Derkaoui:MonopoleIsotropy}
J.~Derkaoui {\em et~al.},
\newblock Astropart. Phys. {\bf 9}, 173 (1998).

\bibitem{Lundberg:Photonics}
J.~Lundberg {\em et~al.},
\newblock Nucl. Instrum. Meth. {\bf A581}, 619 (2007), astro-ph/0702108.

\bibitem{IceCube:IceModel_AHA}
M.~Ackermann {\em et~al.},
\newblock Journal of Geophysical Research D {\bf 111}, 13203 (2006).

\bibitem{IceCube:SPICEMie}
IceCube Collaboration, M.~Aartsen {\em et~al.},
\newblock Nucl. Instrum. Meth. {\bf A711}, 73 (2013), arXiv/1301.5361.

\bibitem{CORSIKA}
J.~Capdevielle {\em et~al.},
\newblock {\em {The Karlsruhe Extensive Air Shower Simulation Code CORSIKA
  }}KfK (Series) (Kernforschungszentrum, Institut f{\"u}r Kernphysik, 1992).

\bibitem{Stanev:SIBYLL}
R.~Fletcher, T.~Gaisser, P.~Lipari, and T.~Stanev,
\newblock Phys. Rev. {\bf D50}, 5710 (1994).

\bibitem{Hoerandel:Flux}
J.~R. Hoerandel,
\newblock Astropart. Phys. {\bf 19}, 193 (2003), astro-ph/0210453.

\bibitem{Gluesenkamp:Monopoles}
T.~Gl\"usenkamp,
\newblock On the {D}etection of {S}ubrelativistic {M}agnetic {M}onopoles with
  the {IceCube} {N}eutrino {O}bservatory,
\newblock Diplomarbeit, RWTH Aachen, 2010.

\bibitem{Ahrens:AMANDA_Reconstruction}
AMANDA Collaboration, J.~Ahrens {\em et~al.},
\newblock Nucl. Instrum. Meth. {\bf A524}, 169 (2004), astro-ph/0407044.

\bibitem{Hill:MRF}
G.~C. Hill and K.~Rawlins,
\newblock Astropart. Phys. {\bf 19}, 393 (2003), astro-ph/0209350.

\bibitem{icecube:muonfilter}
IceCube Collaboration, M.~Aartsen {\em et~al.},
\newblock (2013), arXiv/1307.6669.

\bibitem{icecube:cascade_filter}
IceCube Collaboration, M.~Aartsen {\em et~al.},
\newblock (2013), arXiv/1312.0104.

\bibitem{Hocker:TMVA}
A.~Hocker {\em et~al.},
\newblock PoS {\bf ACAT}, 040 (2007), physics/0703039.

\bibitem{BDT}
P.~Byron, H.~Yang, and J.~Zhub,
\newblock Boosted decision trees, a powerful event classifier,
\newblock in {\em Statistical Problems in Particle Physics, Astrophysics and
  Cosmology: proceedings of PHYSTAT05} Vol.~40, p. 139, Imperial College Pr,
  2006.

\bibitem{RolkeLimits}
W.~A. Rolke, A.~M. Lopez, and J.~Conrad,
\newblock Nucl. Instrum. Meth. {\bf A551}, 493 (2005), physics/0403059.

\bibitem{EarlyUniverse:Kolb}
E.~Kolb and M.~Turner,
\newblock {\em The Early Universe,\;}Frontiers in Physics (Addison-Wesley
  Longman, Incorporated, 1990).

\bibitem{indirectSKlimits}
Super-Kamiokande Collaboration, K.~Ueno {\em et~al.},
\newblock Astropart. Phys. {\bf 36}, 131 (2012), arXiv/1203.0940.

\bibitem{Harvey:Bounds_neutronstars}
J.~A. Harvey,
\newblock Nucl. Phys. B {\bf 236}, 255 (1984).

\bibitem{Freese:Bounds_whitedwarfs}
K.~Freese and E.~Krasteva,
\newblock Phys. Rev. D {\bf 59}, 063007 (1999), astro-ph/9804148.

\bibitem{Arafune:Bounds_jovianplanets}
J.~Arafune, M.~Fukugita, and S.~Yanagita,
\newblock Phys. Rev. D {\bf 32}, 2586 (1985).

\bibitem{IceCube:ICRC2013_slowMPs}
IceCube Collaboration, M.~Aartsen {\em et~al.},
\newblock p.~25 (2013), arXiv/1309.7007.

\bibitem{Zierke:Masterarbeit}
S.~Zierke,
\newblock {V}erbesserung von {R}ekonstruktions- und {Datenselektions-Methoden}
  {f\"ur} die {Messung} subrelativistischer magnetischer {Monopole} mit
  {IceCube},
\newblock Master thesis, RWTH Aachen, 2013.

\end{thebibliography}

\end{document}